\documentclass[acmlarge,screen]{acmart}

\AtBeginDocument{%
  \providecommand\BibTeX{{%
    \normalfont B\kern-0.5em{\scshape i\kern-0.25em b}\kern-0.8em\TeX}}}

\setcopyright{acmlicensed}
\copyrightyear{2025}
\acmYear{2025}
\acmDOI{XXXXXXX.XXXXXXX}

\acmJournal{TIST} 

\PassOptionsToPackage{table,x11names}{xcolor}

\newcommand{\dquotes}[1]{``#1''}
\newcommand{\squotes}[1]{`#1'}

\usepackage{tcolorbox}
\usepackage{tabularx}
\usepackage[table]{xcolor} 
\usepackage{multirow}
\usepackage{booktabs}
\usepackage{makecell}
\usepackage{graphicx}
\usepackage{caption}
\usepackage{subcaption}

\usepackage{tabularx}
\usepackage{booktabs}
\usepackage{hyperref}
\usepackage{tikz}
\usetikzlibrary{shapes.geometric, arrows, positioning}

\usetikzlibrary{shapes.geometric, arrows.meta, chains, positioning, quotes}

\usepackage{xcolor}

\definecolor{lightgray}{gray}{0.9}
\definecolor{customblue}{rgb}{0, 0, 0} 

\usepackage[framemethod=TikZ]{mdframed}

\begin{document}

\title[CFaiRLLM: Consumer Fairness Evaluation in LLM Recommender Systems]{CFaiRLLM: Consumer Fairness Evaluation in Large-Language Model Recommender System}

\author{Yashar Deldjoo}
\email{deldjooy@acm.org}
\orcid{0000-0002-6767-358X}
\affiliation{%
  \institution{Politecnico di Bari}
  \streetaddress{ Via Amendola 126/B}
  \city{Bari}
  \country{Italy}
  \postcode{70126}
}
\author{Tommaso di Noia}
\email{deldjooy@acm.org}
\orcid{0000-0002-6767-358X}
\affiliation{%
  \institution{Politecnico di Bari}
  \streetaddress{ Via Amendola 126/B}
  \city{Bari}
  \country{Italy}
  \postcode{70126}
}

\renewcommand{\shortauthors}{Yashar Deldjoo, and Tommaso di Noia}

\begin{abstract}

\textcolor{customblue}{This work takes a critical stance on previous studies concerning fairness evaluation in Large Language Model (LLM)-based recommender systems, which have primarily assessed consumer fairness by comparing recommendation lists generated with and without sensitive user attributes. Such approaches implicitly treat discrepancies in recommended items as biases, overlooking whether these changes might stem from genuine personalization aligned with true preferences of users. Moreover, these earlier studies typically address single sensitive attributes in isolation, neglecting the complex interplay of intersectional identities. In response to these shortcomings, we introduce \textbf{CFaiRLLM,} an enhanced evaluation framework that not only incorporates \textit{true preference alignment} but also rigorously examines \textit{intersectional fairness} by considering overlapping sensitive attributes. Additionally, CFaiRLLM introduces diverse user profile sampling strategies—\textit{random}, \textit{top-rated}, and \textit{recency-focused}—to better understand the impact of profile generation fed to LLMs in light of inherent token limitations in these systems. Given that fairness depends on accurately understanding users' tastes and preferences, these strategies provide a more realistic assessment of fairness within RecLLMs.}

\textcolor{customblue}{To validate the efficacy of CFaiRLLM, we conducted extensive experiments using~\texttt{MovieLens} and~\texttt{LastFM} datasets, applying various sampling strategies and sensitive attribute configurations. The evaluation metrics include both item similarity measures and true preference alignment considering both hit and ranking (Jaccard Similarity and PRAG), thereby conducting a multifaceted analysis of recommendation fairness. The results demonstrated that true preference alignment offers a more personalized and fair assessment compared to similarity-based measures, revealing significant disparities when sensitive and intersectional attributes are incorporated. Notably, our study finds that intersectional attributes amplify fairness gaps more prominently, especially in less structured domains such as music recommendations in LastFM. These findings suggest that future fairness evaluations in RecLLMs should incorporate true preference alignment to ensure equitable and genuinely personalized recommendations.}

\end{abstract}






\maketitle

\section{Introduction}
\label{sec:into}

Recently, recommender systems driven by large language models (RecLLM), such as ChatGPT, have received substantial attention from the research community, becoming an important research area in the fields of Information Retrieval (IR) and Recommender Systems (RS). These sophisticated neural architectures utilize billion-scale parameters trained through supervised and semi-supervised methods on extensive internet data. They have shown significant potential in various sectors and tasks, including but not limited to healthcare~\cite{nazary2023chatgpt,jin2024health,nazary2024xai4llm}, finance~\cite{dowling2023chatgpt,wu2023bloomberggpt}, conversational assistants~\cite{biancofiore2024interactive,liao2023proactive,he2023large}, and many more, see e.g., recent surveys~\cite{deldjoo2024review,deldjoo2024recommendation,zhang2024understanding,chang2023survey} for a good frame of reference. Despite the witnessed benefits of using these systems in top-$k$ recommendation setting~\cite{sanner2023large,he2023large}, there are rising concerns about their inherent biases~\cite{deldjoo2023fairnessgpt,deldjoo2023fairness,di2023evaluating}. The vast and unregulated nature of the Internet data used to train Large Language Models (LLMs) raises alarms about possible biases against specific races, genders, popular brands, and other sensitive attributes that could be \textbf{encoded} in these networks. For example, if an LLM is predominantly trained on data from popular e-commerce sites, it might disproportionately recommend products from more recognized brands, overlooking niche or emerging brands. Similarly, biases in language around gender or race could skew recommendations in subtle but impactful ways. Hence, \underline{unchecked} employment of these systems in commercial RS may lead to unfair treatment of minority groups with societal impacts, such as reinforcing existing stereotypes or exacerbating economic disparities. 

Recent studies~\cite{wu2023survey,li2023large,xu2024prompting,deldjoo2024review} highlight that recommender systems can harness Large Language Models (LLMs) in three key ways: (i) as the core recommender, (ii) as a means of data augmentation, incorporating rich semantic representations from textual data, and (iii) as simulators to refine system's predictions. Our work focuses on the first application, wherein an LLM provides personalized recommendations, given a textual query provided by the user, in the form of a \textit{prompt}. In such a case, users indicate their preferences through textual prompts, such as requesting movie suggestions based on their recent views, e.g., \textit{"Based on the movies I have recently watched: Blade (1998) (Genres: Action\textbar Adventure\textbar Horror), and Four Weddings and a Funeral (1994) (Genres: Comedy\textbar Romance), please provide me with 3 movie recommendations."} Utilizing their deep understanding of context, user preferences, and an extensive knowledge base, LLMs can propose relevant movie suggestions. These recommendations are then verified against the existing catalog to ensure availability before being presented to the user.

Building upon our focus on employing LLMs for personalized recommendations, our research particularly emphasizes the \textbf{fairness of RecLLMs}. We can examine the broad literature and taxonomies presented in FairRS from two perspectives: (i) evaluation and (ii) system design. Regarding the former \textit{evaluation perspective}, the literature on FairRS identifies numerous noteworthy dimensions. These include the stakeholder perspective (consumer vs. producer), the nature of benefits being examined (effectiveness or exposure), the level of fairness (individual or group), and the core definition of fairness, among others~\cite{ekstrand2019fairness,deldjoo2021explaining}. This focus contrasts with the \textit{system perspective}, which concentrates on the core recommendation model. In contrast, from a system perspective -- i.e., which core recommendation model is being employed ---, a notable observation in the FairRS literature is that they primarily investigate so-called \textbf{conventional/traditional} models, based on collaborative filtering models such as matrix-factorization (MF), or variations thereon NeuMF, LightGCN, rather than those based on Large Language Models (LLMs). As shown in Table~\ref{tab:fairness_mapping}, only a limited number of studies~\cite{zhang2023chatgpt,li2023large,di2023evaluating,deldjoo2024understanding} have explored the fairness aspects of RecLLMs, not least due to their relative novelty and recent development.

\begin{figure*}
    \centering
    \includegraphics[width = 0.95\linewidth]{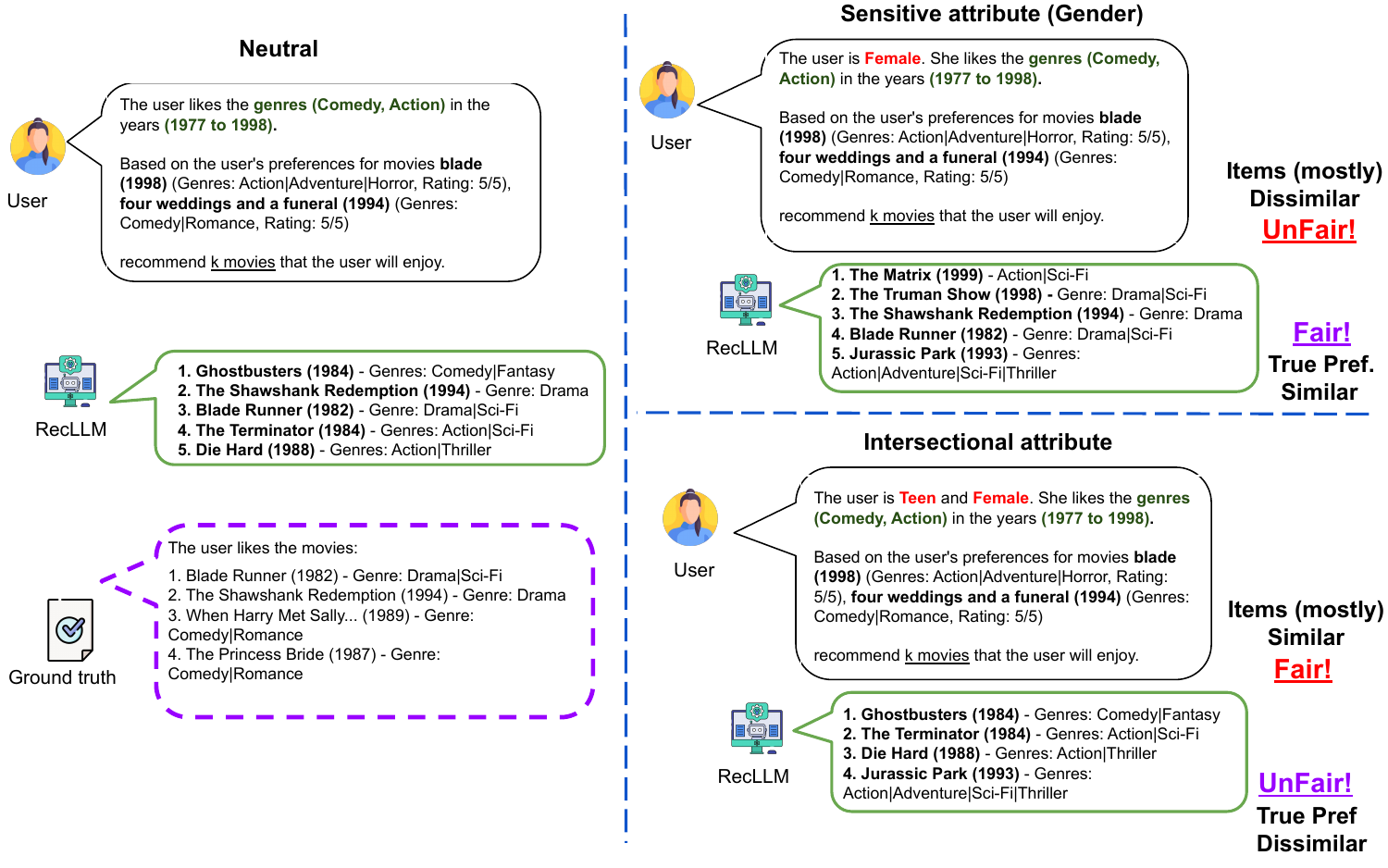}
    \caption{In the left figure, we showcase CFaiRLLM's fairness evaluation in movie recommendations, comparing recommendation similarity across sensitive (gender, age) and intersectional attributes to a neutral standard, emphasizing user preferences. Our aim is equity, ensuring that sensitive attribute recommendations align with neutral benchmarks. The right details the sensitive attributes explored.}
        \label{fig:fairness_evaluation}

\end{figure*}

The research work by~\citet{li2023large} focuses on issues of producer fairness and personalization in RecLLMs, specifically examining ChatGPT in the context of news recommendations. Other recent work such as~\cite{deldjoo2024understanding} focuses on the producer's perspective, scrutinizing the impact of prompt engineering on system personalization, item fairness, and the tendency of GPT-based RecLLMs to favor more recent, post-2000 movie recommendations. The research carried out by~\citet{zhang2023chatgpt} forms the basis of our research, in which the authors study the fairness of zero-shot GPT recommendations. Their work introduces an evaluation framework called FaiRLLM, designed to assess fairness in Large Language Model recommendations (RecLLM), particularly focused on the consumer side. This framework provides specialized evaluation metrics and datasets for evaluating fairness across various sensitive user attributes in different recommendation scenarios, such as music and movies. Their evaluation of ChatGPT using FaiRLLM reveals notable biases toward certain sensitive attributes, underscoring the need for further investigation and mitigation of these biases.

Our study advances the discourse in FairRS/FairLLM by introducing an enhanced framework for assessing  \dquotes{\textbf{Consumer Fairness in RecLLMs}}, with a particular emphasis on two aspects: (i) the definition of unfairness (whether through similarity alignment or true preference alignment), and (ii) the granularity of groups (considering both individual and intersectional prompts). We also study the \dquotes{user-profile construction} strategies, as well as the \textit{scope of recommendations} prompted in the aforementioned study.

Our research builds on and meticulously refines the foundational framework suggested by ~\citet{zhang2023chatgpt}, enhancing its evaluation foundation and application in several key ways. Our approach notably expands upon the methodology proposed by these researchers, which involves comparing the recommendations generated by LLMs under neutral conditions to those produced when sensitive attributes are revealed. For example, when a user requests movie recommendations without disclosing sensitive attributes (e.g., \dquotes{\textit{Please recommend movies you think I would enjoy.}}), the RecLLM might offer a diverse selection based on the user's past interactions. However, if the user specifies a sensitive attribute such as age or gender (e.g., \dquotes{\textit{As a woman interested in movies, please recommend...}}), the existing FairLLM framework would compare the similarity of this personalized recommendation list with the original, using such discrepancies to flag potential consumer unfairness. At first glance, this approach aligns with a widely accepted notion of fairness, positing that system performance should remain consistent and must not vary based on sensitive attributes. Building upon this principle, the authors applied their framework to both music and movie recommendations and identified numerous biases.

However, this approach presents certain limitations. First, the most crucial limitation is the presumption that difference in recommendations inherently means unfairness, overlooking the possibility that such differences could simply represent \dquotes{personalization}, which is not inherently negative. Their methodology equates the disparity in recommendation lists with fairness issues without scrutinizing the \textit{preference alignment gap} -- that is, whether the recommendations accurately reflect the user's actual preferences. Second, the framework gauges fairness based on a \textit{single sensitive attribute class} (e.g., gender or age group) in isolation, and overlooks the complexity of overlapping identities. For example, a user known for favoring action movies receives a list of high-octane films like \dquotes{Mad Max: Fury Road} and \dquotes{John Wick} in response to a generic prompt. However, when the prompt includes intersectional features, such as \dquotes{\textit{I am a middle-aged woman looking for good movies to watch,}} the system's recommendations shift toward stereo-typically \textbf{gender-} and \textbf{age-associated} films such as \dquotes{Under the Tuscan Sun} and \dquotes{Eat Pray Love.} If the user however adds only her gender to the prompt: \dquotes{\textit{I am a woman interested in great movies.}} The recommendations shift significantly, leaning towards movies like \dquotes{Little Women,} \dquotes{Pride and Prejudice,} and \dquotes{The Help,} reflecting a \textbf{stereotypical assumption} about gender-specific preferences. Overall, these examples and ideas reveal a preference alignment gap where the personalization is skewed by demographic assumptions rather than the user's demonstrated taste for movies.

\noindent \textbf{Example.} In our CFaiRLLM framework depicted in Figure~\ref{fig:fairness_evaluation}, we conduct an in-depth examination of recommendation fairness by contrasting the impact of single sensitive with intersectional prompts. Let us consider a user profile with a history of enjoying \textit{Comedy} and \textit{Action} movies from the period of 1977 to 1998. \vspace{2mm}

\begin{itemize}
    \item \textit{Gender-based Prompt.} When the \textbf{user's gender} is considered, RecLLM modifies the recommendation set. For a female user, the system might recommend \textit{The Matrix} (1999), \textit{The Truman Show (1998)},  \textit{The Shawshank Redemption (1994)}, \textit{Blade Runner (1982)} , and \textit{Jurassic Park (1993)} which despite being high-quality movies within the action and drama genres, they are (mostly) different from those given without considering gender. This change is deemed unfair according to \cite{zhang2023chatgpt} for relying on gender-based assumptions rather than actual interests. In contrast, we argue this recommendation is fair because both gender-specific and neutral prompts yield only two movies, \textit{The Shawshank Redemption (1994)} and \textit{Blade Runner (1982)}, that aligns with ground truth. In other words, both gender-specific and neutral prompts share one common \underline{accurate} recommendation (\textit{Blade Runner}), indicating no unfair advantage (here benefit is equal to proving better recommendation in terms of relevance/quality).
    \item \textit{Intersectional-Prompt.} When a user's identity includes both \textbf{Teen} and \textbf{Female} sensitive attributes, RecLLM changes its recommendations to reflect this intersectionality. The system suggests movies such as  \textit{Ghostbusters (1984),}  \textit{The Terminator (1984)}, and \textit{Die Hard (1988)}, and \textit{Jurassic Park (1993)}, which closely resemble those from a neutral recommendation. However, the recommendations tailored to these sensitive attributes do not match any items in the ground truth, unlike the neutral recommendations, which include two items from the ground truth. Consequently, despite the previous approach, in our work this system is considered unfair, as it appears to favor the neutral recommendations by providing a more accurate reflection of user preferences.

\end{itemize}

 The core insight behind our framework evaluates fairness by examining if the recommendations are truly personalized or if they are biased by stereotypes associated with the sensitive attributes. For example, if the intersectional approach yields recommendations for movies, which may align with stereotypical views of teen girls' preferences, the framework would flag this as potentially unfair. The fairness is adjudged by the degree to which the recommendations align with the user's actual, demonstrated preferences—such as their affinity for action comedies from the late 20th century—regardless of their gender or age. \vspace{1.2mm}

\subsection{Contributions.} Our work offers the following list of contributions:

\begin{enumerate}
    \item \textbf{Introduction of an Enhanced Evaluation Framework for Consumer Fairness in RecLLMs.} Our work improves the research on FairRS in RecLLMs by proposing, \textbf{CfaiRLLM}, a more detailed framework that evaluates consumer fairness with an emphasis on the \textit{true alignment of recommendations} when measuring benefits in RS. This includes the analysis of \textit{intersectionality}, or more precisely \textit{intersectional prompts}, encompassing overlapping groups. 

    \vspace{1mm}

\item \textbf{Investigation of Intersectional Prompts in RecLLMs.} This work highlights the role of overlapping groups in group fairness research specifically within RecLLMs. It  studies how combining multiple sensitive attributes (e.g., gender and age) with intersectional prompts affects recommendation fairness. \\

\item  \textbf{Enhanced Understanding of Unfairness Through User Profile Sampling Strategies.} Our work proposes a suite of different user profile sampling strategies (\squotes{random}, \squotes{top-rated}, \squotes{recent}), with the goal to study how these strategies influence the fairness of recommendations. This contribution is essential within RecLLM research for developing more equitable recommender systems that mitigate bias.

Our work designs and studies the impact of the following strategies through the course of experiments.
    \begin{itemize}
       \item \textit{Random Sampling}: Examining the fairness and relevance of recommendations when a random set of movies from the user's history is used to generate new recommendations.
   \item  \textit{Top-Rated Sampling}: Analyzing how using the user's top-rated movies to generate recommendations affects the alignment with their preferences and potential biases.
    \item \textit{Recent Sampling}: Investigate the impact of prioritizing movies most recently watched or rated by the user on the fairness and relevance of recommendations.

   \end{itemize}
   
   \item \textbf{Comparison with Existing Work.} The research builds upon and refines the foundational framework suggested by \citet{zhang2023chatgpt}, enhancing its evaluation foundation and application. It provides a comparative analysis that not only acknowledges the contributions of prior work but also identifies and addresses its limitations.

\end{enumerate}

   \vspace{0.8mm}

In essence, the fairness of RecLLM recommendations is determined not just by the presence of similarity but by the depth of alignment with the users' true preferences. Our framework seeks to ensure that  recommendations are fair and personalized, moving beyond stereotypes. \\

\noindent \textbf{Note.} \textcolor{customblue}{As illustrated in Figure \ref{fig:fairness_evaluation}, one might question whether simply designing an LLM-based recommender system to strip gender (or other sensitive attribute) terms from queries would effectively solve the problem. To answer this quesry, we provide the following viewpoints:}

\begin{itemize}
\item \textcolor{customblue}{Users may not always use explicit phrases to convey their identity, but their interactions and preferences can implicitly reflect sensitive attributes. By incorporating sensitive attributes in our prompts, we aim to simulate scenarios where user identity influences recommendations, either explicitly or implicitly.}

\item \textcolor{customblue}{In real-world applications, users exhibit a wide range of behaviors in how they express their preferences. Some users might naturally include identity-related information in their queries to receive more tailored recommendations. Our framework accounts for this diversity by evaluating how such expressions can surface or mitigate biases.}

\end{itemize}

\textcolor{customblue}{Overall, even if explicit mentions of sensitive attributes are rare, biases can still permeate through the data and influence recommendations in subtle ways. These considerations aim to ensure that our CFaiRLLM framework provides a broader assessment of fairness, taking into account both overt and nuanced impact of sensitive attributes on recommendation outcomes.}

\section{Related work}
\label{sec:related_work}

In this section, we briefly review some related work on
recommendation systems and LLM techniques.

\subsection{Fairness in Recommender Systems}

In examining the landscape of FairRS research, we can categorize the literature along several dimensions: the \textit{stakeholder focus}, the \textit{core recommender system model}, the \textit{dynamics of fairness evaluation}, and the granularity of fairness with respect to group and individual distinctions~\cite{deldjoo2023fairness}. Table~\ref{tab:fairness_mapping} provides an overview of how different papers are categorized by stakeholder focus—consumer or producer— and RS models, whether traditional or employing recent RecLLM advances. This table further elucidates the attributes used to operationalize fairness considerations from both consumer and producer standpoints, underlining the relative scarcity of research focusing on RecLLMs within the FairRS domain. Table~\ref{tab:granularity_longitudinal} instead provide an additional discourse by underscoring the technical nuances of fairness evaluation, differentiating between static and dynamic methodologies. It highlights the dominance of group and static evaluations in the current literature, pointing to potential areas for further investigation and development.
\vspace{3mm}

\subsubsection{Core RS models and Stakeholder.} In the current landscape of recommender systems (RS), in particular, on FairRS, we observe a clear division between traditional models and those enhanced by recommendation-centric large language models (RecLLM).  Traditional RS principally operates on collaborative filtering (CF) mechanisms, potentially supplemented by auxiliary user and item side information. These systems have been the subject of numerous studies aimed at developing benchmarks for fairness evaluation and strategies for bias mitigation, as studied in depth in ~\cite{deldjoo2023fairness,ekstrand2019fairness}. Traditional RS models serve as the backbone of recommendation systems, albeit without the intricate natural language processing capabilities endowed by large language models (LLMs). The \squotes{RecLLM} paradigm represents an innovative frontier in AI and RS research~\cite{kattnig2024assessing,kattnig2024assessing}. Here they refer to models that integrate complex NLP methods, such as those derived from GPT-like architectures, into the recommendation process. While this integration promises enhanced personalization and a refined recommendation experience by comprehending user nuances, it simultaneously poses the risk of inheriting and perpetuating biases present in the extensive and unfiltered data used for LLM training. Concerns are particularly pronounced regarding biases related to race, gender, and brand recognition, which could result in unbalanced exposure for emerging entities or products. Consequently, a fundamental aspect of our ongoing research is the precise quantification of these biases, setting the stage for the formulation of strategies that can effectively neutralize the inadvertent propagation of these biases through (advanced) RecLLM systems.

\begin{table}[htbp]
\centering
\caption{Mapping of research papers to core models and stakeholder fairness. \checkmark\textbf{**} positions our work within the FairRS literature.}
\label{tab:fairness_mapping}
\begin{tabularx}{\textwidth}{@{}cXcc cccc@{}}
\toprule
\textbf{Category} & \textbf{attributes} & \multicolumn{2}{c}{\textbf{Core RS Model}} & \phantom{a} & \multicolumn{3}{c}{\textbf{Stakeholder}} \\
\cmidrule(lr){3-4} \cmidrule(lr){6-8}
                  &                   & \textbf{\small Traditional}  & \textbf{\small RecLLM} & & \textbf{\small Consumer} & \textbf{\small Producer} & \textbf{\small CP Fairness} \\
\midrule
\small{\cite{li2021user,hao2021pareto,xiao2020enhanced}}                  & \small{consumer activity}       & \checkmark  &                 & & \checkmark & & \\  \cmidrule(lr){1-8}

\small{\cite{deldjoo2021flexible,deldjoo2021explaining,wu2021fairness,weydemann2019defining,farnadi2018fairness}}                   & \multirow{2}{*}{\small consumer demographics}          & \checkmark  &                 & & \checkmark & & \\

\small{\cite{shen2023towards,zhang2023chatgpt}} 
                &             &   &\checkmark \checkmark\textbf{*}                 & & \checkmark\checkmark\textbf{*}   & & \\ 
                \cmidrule(lr){1-8}
\small{\cite{suhr2021does,gomez2021winner}}                 & \small{consumer merits}       & \checkmark  &                 & & \checkmark & & \\  \cmidrule(lr){1-8}
           \small{\cite{wan2020addressing,lin2021mitigating}}                 & \small{other consumer attributes}       & \checkmark  &                 & & \checkmark & & \\  \midrule

\small{\cite{dong2021user,da2021exploiting,ge2021towards,zhu2021fairness}}                   & \multirow{2}{*}{\small{producer/item popularity}}          & \checkmark  &                 & &  & \checkmark& \\

\small{\cite{deldjoo2024understanding,li2023preliminary}} 
                &             &   &\checkmark                 & &  &\checkmark & \\ \cmidrule(lr){1-8}
\small{\cite{kirnap2021estimation,boratto2021interplay}}                & \small{producer demographics}       & \checkmark  &                 & & &\checkmark & \\  \midrule

\small{\cite{deldjoo2021flexible,burke2018balanced,liu2020balancing,shakespeare2020exploring}}    & \small{price/brand/location}       & \checkmark  &                 & & &\checkmark & \\  \midrule
\small{\cite{naghiaei2022cpfair,rahmani2022unfairness,chakraborty2017fair,patro2020fairrec,wu2021tfrom,do2021two}} & \small{variate CP attributes}       & \checkmark  &                 & & & &\checkmark \\

\bottomrule
\end{tabularx}
\end{table}

In summary:
\begin{itemize}
    \item \textbf{Traditional RS.}  In \squotes{traditional RS}, the focus frequently lies on~\textit{collaborative filtering (CF)} algorithms that utilize~\textit{historical datasets} within a~\textit{train-predict} paradigm, occasionally supplemented by user and item metadata. Although these systems operate effectively, they lack the advanced natural language processing (NLP) capabilities seen in large language models (LLMs). Research work in this area are dedicated to enhancing frameworks for evaluating fairness and formulating strategies to mitigate biases inherent in these models. \vspace{1mm}
    
    For example,~\citet{hao2021pareto}  tackle the issue of unfair discrimination by CF due to imbalanced data, proposing a multi-objective optimization approach that seeks a Pareto optimal solution to balance subgroup performance without sacrificing overall accuracy.~\citet{farnadi2018fairness} address inherent biases by introducing a hybrid fairness-aware recommender system that merges multiple similarity measures and demographic information to mitigate recommendation biases.~\citet{naghiaei2022cpfair}  highlight the two-sided nature of recommender systems and present a re-ranking approach that integrates fairness constraints for both consumers and producers, showcasing the algorithm's ability to improve fairness without diminishing recommendation quality. Lastly,~\citet{wan2020addressing} investigate the bias induced by marketing strategies in CF systems and propose a framework that enhances fairness across different market segments, achieving more equitable recommendation performance. \vspace{2mm}

    \item \textbf{RecLLM.} These model address the integration of language models with RS, signifying a  shift towards using NLP techniques to refine the accuracy and pertinence of recommendations. Research in this domain is not limited to the application of LLMs for classical top-$k$ recommendation tasks but also extends to applications such as conversational recommendation systems, personalized explanation generation, and multi-modal recommendation scenarios. \vspace{2mm}

    For instance,~\citet{shen2023towards} examine the unintended biases in language model-driven Conversational Recommendation Systems (CRSs), showing how biases can influence the category and price range of recommendations, and offer mitigation strategies that preserve recommendation quality. \citet{li2023preliminary} study the application of ChatGPT in personalized news recommendation task and find that the system is sensitive to input phrasing and signal the challenges in achieving provider fairness and fake news detection. They suggest that ongoing, dynamic evaluation of ChatGPT's recommendations is crucial for understanding and improving its performance in real-world tasks. Another study by~\citet{zhang2023chatgpt} introduces the FaiRLLM benchmark, as pointed out earlier, specifically designed to evaluate the fairness of recommendations produced by RecLLM systems, highlighting the biases these models exhibit against certain sensitive user attributes in music and movie recommendations. Recently, Deldjoo et al.~\cite{deldjoo2024understanding} investigate prompt design strategies within ChatGPT-based RecLMMs and assess their effect on recommendation quality, and provider fairness. They find that assigning system role can mitigate popularity bias and enhance fairness, suggesting that combining these strategies with personalized models could lead to a more balanced recommendation experience. These examples underscore the evolving nature of RS technology and the importance of considering biases and fairness in the development of RecLLM systems.

\end{itemize}

In general, the comparison of these two RS methodologies in Table \ref{tab:fairness_mapping} highlights a critical moment in the evolution of RS, where the quest for superior personalized recommendations must be meticulously weighed against the essential need for fairness and other harms in every facet of the recommendation process. \vspace{2mm}

\subsubsection{Stakeholder Considerations.} Earlier discussions have established the importance of market orientation in classifying the corpus of group fairness research—whether they address concerns from the consumer perspective, the provider's angle, or a combination of both. In delineating these categories, literature often focuses on certain sensitive attributes such as consumer demographics (including age and gender) and producer-related attributes such as item popularity.

\begin{itemize}
    \item \textbf{Consumer Fairness.} Research in this area aims to ensure equitable recommendations for consumers, where fairness is typically measure based on the relevance (or effectiveness) of recommendations for user groups, e.g., demographic groups. Typically, as illustrated in Table \ref{tab:fairness_mapping}, consumer level of activity (e.g., active vs. inactive users), demographics, or other metrics (e.g., education) are utilized to identify protected groups.
    
\item \textbf{Producer Fairness.} This aims to achieve fairness for content or product creators within the recommender system. Fairness could be measured at the item level (e.g., popularity of items) or the producer level (such as artists, authors, brands), with popularity or recognition of artists/brands as examples. In technical terms, a producer can be seen as a higher-level grouping of items. Several attributes have been used, including popularity, demographics, and price/brand/location.

\item \textbf{CP Fairness.} This encompasses research that considers both consumer and producer fairness, endeavoring to achieve a balanced approach.
\end{itemize}

\vspace{4mm}

\noindent \textbf{Positioning the current work.} While fairness in traditional recommendation systems is well-established, we observe a scarcity of research on fairness in LLMs. The present study seeks to address this gap by focusing on evaluating fairness and biases within RecLLMs. Our attention is particularly drawn to the consumer aspect of RecLLMs, building upon and refining previous works, especially since significant research has already been conducted on demographics from the consumer perspective. However, we also aim to propose a similar framework for the producer side and eventually explore a combined approach that integrates both consumer and producer perspectives. Our work also addresses the static aspects of fairness in recommender systems, as shown in Table~\ref{tab:granularity_longitudinal}. The purpose of presenting this table is to highlight the ample opportunities for further research in this field.

\begin{table}[htbp]
\centering
\caption{Research papers classification by granularity and longitudinal criteria. \textbf{*} positions our work.}
\label{tab:granularity_longitudinal}
\begin{tabular}{l p{0.4\textwidth} p{0.4\textwidth}}
\toprule
 & \textbf{Static} & \textbf{Dynamic} \\
\midrule
\textbf{Individual} & \cite{burke2018balanced,liu2019advertisement,zhang2023chatgpt} & \cite{zhu2021fairness,do2021two}  \\
\textbf{Group} & \cite{deldjoo2024understanding,hao2021pareto,xiao2020enhanced,deldjoo2021flexible,deldjoo2021explaining,wu2021fairness,shen2023towards}* & \cite{ge2021towards,chakraborty2017fair,do2021two, burke2018balanced,liu2020balancing} \\
\bottomrule
\end{tabular}
\end{table}


\subsection{Leveraging Pre-trained LMs and Prompting for Recommender Systems}

The integration of natural language processing (NLP) techniques within RS, underscores the major role of LLMs in enhancing recommendation accuracy through deep semantic understanding. For instance,~\citet{hou2022towards} utilize natural language descriptions and tags as inputs into LLMs to create user representations for more effective recommendations. This contrasts with the narrative-driven recommendations~\cite{bogers2017defining} that rely on verbose descriptions of specific contextual needs.

Regarding the evolution of prompting strategies, initial attempts often employed few-shot learning~\cite{brown2020language}, guiding LLMs using exemplary cases to refine task-specific outcomes. Through the progress of prompt learning, tasks are adapted to align with LLM capabilities, rather than adapting LLMs to tasks,  employing either discrete or continuous/soft prompts to improve performance across various tasks. This strategy has demonstrated effectiveness across a range of tasks, including recommendation tasks.

At the core of these advances lies the personalization of LLMs for recommendation purposes. The P5 framework~\cite{geng2022recommendation} and its iterations, such as OpenP5~\cite{xu2023openp5}, showcase the integration of multiple recommendation tasks into a unified LLM framework using personalized prompts. This approach reformulates recommendation tasks as sequence-to-sequence generation problems, showing the adaptability of LLMs to various recommendation contexts and emphasizing the importance of capturing user intent and personalized needs. Furthermore, exploring prompt transfer techniques, such as SPoT~\cite{vu2021spot}  and ATTEMPT~\cite{asai2022attempt}, represents a major step in applying the learned knowledge from source tasks to target recommendation tasks. These methodologies, together with knowledge distillation techniques, contribute significantly to the development of more efficient and effective LLM-based recommendation models. They underscore the potential for intra-task prompt distillation and cross-task prompt transfer, enhancing the efficiency and effectiveness of LLM-based recommendation models.

In sum, the integration of LLMs into recommender systems represents a paradigm shift towards leveraging advanced NLP and innovative prompting strategies for delivering highly personalized and contextually rich recommendations. These developments promise to reshape the landscape of recommender systems, making them more adaptable, intuitive, and user-centric.

\section{Proposed Evaluation Framework}

The integration of Large Language Models (LLMs) into recommendation systems (RecLLMs) has underscored the critical need for a thorough evaluation of fairness in the recommendations provided to users. We introduce  \textbf{CFairLLM}, a rigorous framework specifically designed to assess the fairness of RecLLMs from a consumer perspective. This framework is an extension and enhancement of the FaiRLLM benchmark, originally proposed by~\citet{zhang2023chatgpt}. It refines the conceptualization of fairness and systematically addresses the limitations inherent in the FaiRLLM framework.

\subsection{CFairLLM: Consumer Fairness Evaluation RecLLM}
\subsubsection{Fairness Definition}
\label{subsec:def}

The concept of consumer fairness in our CFairLLM framework is predicated on the impacts of sensitive attributes on the outcomes produced by RecLLMs. We use the original definition proposed by~\cite{zhang2023chatgpt} maintaining its original phrasing and intent. \vspace{1mm}

According to~\citet{zhang2023chatgpt}: \vspace{0.5mm}

\begin{description}
   \item  Given a sensitive attribute (e.g., gender) of users, fairness of RecLLM on the consumer-side could be  defined as  \dquotes{\textit{the absence of any prejudice or favoritism toward user groups with specific values (e.g., male vs. female) of the sensitive attribute when generating recommendations \underline{without} using such sensitive information.}} 
\end{description}

This definition essentially emphasizes the importance of treating user groups equally in the process of generating recommendations, regardless of their values for sensitive attributes.

\subsubsection{Limitations and Our Contributions.}
Our work builds on the established definition to highlight that prejudice originates from the nature and distribution of benefits between user groups. We propose to evaluate fairness in terms of the alignment between recommendations and users' actual preferences. Unlike previous studies, such as that of~\citet {zhang2023chatgpt}, which assess fairness by comparing the results of recommenders with and without sensitive data, we argue that fairness should be measured by the consistency of the benefits across different user groups, defined by their true preferences.

In other words, the approach by~\cite{zhang2023chatgpt}, which focuses on the similarity between the ranking lists, might not capture the full picture. Differences in recommendations can arise from personalization or stereotypes that affect fairness. Our approach emphasizes understanding users' genuine preferences to accurately assess fairness, moving beyond~\underline{mere list comparison} to consider the actual benefits to users.

\begin{tcolorbox}[colback=gray!5!white,colframe=gray!75!black]
  \textbf{Example 1.} \vspace{0.5mm}
Imagine a movie streaming service designed to suggest films to its platform users, recommending romantic or classical movies to females and action or sci-fi movies of a more recent vintage to males, based on the stereotype that gender predicts movie taste. This stereotype can misalign with the true preferences of users, as there may be a noteworthy number of females who like recent sci-fi movies. The recommendation system, in this case, is unfair because it aligns its recommendations with stereotypes rather than users' true preferences. An ideal, fair system would align its recommendations with the actual, diverse tastes of its users, regardless of their gender.

\end{tcolorbox}

Furthermore, our framework introduces a nuanced consideration of \dquotes{intersectional fairness,}~\cite{yang2020fairness,deldjoo2023fairness} recognizing that individuals may have multiple overlapping identities that can influence recommendation outcomes. For example, a prompt that includes both \squotes{gender} and \squotes{age}, such as \dquotes{\textit{I am a Young Adult Woman, based on movies I watched, recommend me $k$ movies that I like.}}, requires a response that accounts for this intersection of attributes (i.e., gender \squotes{woman}, and age \squotes{Young}), rather than focusing on a singular demographic attribute (e.g., gender alone).

\begin{tcolorbox}[colback=gray!5!white,colframe=gray!75!black]
\textbf{Example 2.} \vspace{0.5mm}
Consider now an enhanced movie streaming service that integrates \textit{intersectional} considerations into its recommendation algorithms. This service understands that a user's preferences cannot be accurately predicted by a single demographic attribute, such as gender. Therefore, when a user who identifies as a young adult woman interacts with the platform, the system extends beyond just suggesting romantic movies. Instead, it explores a wider category of genres, including thrillers, documentaries, and science fiction, acknowledging her intersectional identity. This approach tailors recommendations to intersect her age group (young adult) and gender (female) with her user profile.
\end{tcolorbox}

We discuss each of the above dimensions in Section \ref{subsec:def} (prejudice and favoritism) and Section \ref{subsec:indep_vs_inter_groups}
 (intersectionality). For a detailed definition, please refer to Section \ref{subsec:eval_proc}.

\subsection{Definition of Rankers and Benefits}
\label{subsec:def}

For the purpose of fairness evaluation, we define two ranking lists provided by two different Large Language Model Recommendation Models (RecLLMs) to assess fairness and bias within our CFairLLM framework:
\begin{itemize}
    \item \textbf{Neutral Ranking List (\(\mathcal{R}_{m}\))}: This list is generated by a RecLLM that operates without any explicit knowledge of sensitive attributes (such as gender, age, ethnicity, etc.). The aim is to simulate a scenario where recommendations are made purely based on user preferences and interactions, without bias or modification influenced by sensitive demographic factors. This list serves as our baseline for fairness, reflecting the model's unbiased recommendations.
    
    \item \textbf{Sensitive Attribute-Influenced Ranking List (\(\mathcal{R}_{m}^a\))}: Contrary to the neutral list, this ranking is produced by a RecLLM that incorporates sensitive attributes (the sensitive attribute here is denoted by $a$) into its recommendation process. The intention here is to observe how the inclusion of such attributes affects the recommendation outcomes. By comparing this list to the neutral ranking list, we can quantify the impact of sensitive attributes on the fairness of recommendations, identifying potential biases introduced by their consideration.
\end{itemize}

The core concept of fairness within the CFairLLM framework, as discussed in Section \ref{subsec:def}, is based on the we define as \textit{benefit} and \textit{prejudice}. To facilitate analysis, we identify two particular types of benefits, represented by the variable \(\mathcal{B}\), which represent the main metrics for comparison and evaluation of fairness in our framework.

\begin{enumerate}
    \item[(i)] \textbf{Alignment of Information Items} (\(\mathcal{B}_{item}\)): This metric assesses the consistency of recommended items across (\(\mathcal{R}_{m}\), \(\mathcal{R}_{m}^{a}\)), i.e.,
    the neutral and sensitive attribute-influenced ranking lists. The principle here, derived from the work of \cite{zhang2023chatgpt}, posits that the fairness of a recommendation system is compromised if the inclusion of sensitive attributes leads to a significant change in the \textit{composition of the top-$k$ recommendation list}. Essentially, this benefit measures the disparity in recommendations, whether any items are unduly favored or omitted, when sensitive attributes are taken into account, treating such disparities as indicators of potential bias.

    \item[(ii)] \textbf{True Preference Alignment} (\(\mathcal{B}_{pref}\)): The notion of benefit introduced in this study is mainly examined through the lens of \textit{user preference}, rather than simply comparing the similarity of recommendation lists. It assesses the extent to which the recommendations from both ranking lists correspond with the users' genuine tastes and interests. Through this metric, we aim to guarantee that both neutral and sensitive attribute rankers deliver equitable recommendation quality to users, regardless of the users' sensitive attributes. Unlike previous approaches, this measure depends on the ground-truth data of the target user.

\end{enumerate}

By leveraging these benefits, \(\mathcal{B}_{item}\) and \(\mathcal{B}_{pref}\), our framework assesses the impact of sensitive attributes on recommendation systems, seeking to maintain the integrity of recommendations by ensuring they are both reflective of true user preferences and consistent across different user groups. 

\subsection{Independent vs. Intersectional Fairness}
\label{subsec:indep_vs_inter_groups}

Fairness in recommendation systems, or more precisely \textit{group fairness}, could be studied through the nuances lens of the \textit{granularity of the sensitive attributes} considered~\cite{yang2020fairness,deldjoo2022survey}, an aspect less explored in the recomemdner system community. Let us consider $A \in \mathcal{A}$, a set of sensitive attributes (such as gender, age), where each element  $A \in \mathcal{A}$  represents one specific category of attributes, (e.g., gender) with associated values that these attributes can take. 

Our framework acknowledges the complexities of intersectional groups and the nuances involved in multiple identities, balancing two principal approaches to fairness: independent groups, which focus on individual sensitive attributes, and intersectional groups~\cite{yang2020fairness}, which examine the overlapping features and combined effects of these attributes. To formalize:

\begin{enumerate}
    \item \textbf{Independent Groups}: Independent groups are formed based on single sensitive attributes, each with multiple potential values. For a given attribute $A \in \mathcal{A}$ with possible values $\{a_1, a_2, \ldots, a_n\}$ (e.g., gender with values male and female), the independent groups are defined as $\mathcal{G}_{\text{indep}} = \{G_{a_1}, G_{a_2}, \ldots, G_{a_n}\}$, where $G_{a_i}$ includes individuals with the attribute value $a_i$. This approach simplifies the assessment of fairness by focusing on one attribute at a time. For clarity, let us assume:
    \begin{itemize}
        \item Gender as an attribute with values $\{a_1: \text{Male}, a_2: \text{Female}\}$,
        \item Age as a separate attribute with its own set of values $\{b_1: \text{Teen}, b_2: \text{Young}, b_3: \text{Adult}\}$. Here, $b_i$ directly corresponds to specific age ranges, illustrating the framework for categorizing individuals based on singular sensitive attributes.
    \end{itemize}
    
    \item \textbf{Intersectional Groups}: For intersectional analysis, we consider the combinations of values from multiple attributes within $\mathcal{A}$. This is formalized as $\mathcal{A}_{int} = \{a_{i} \cap b_{j} \mid a_{i} \in A, b_{j} \in B, A, B \in \mathcal{A}, A \neq B\}$, leading to intersectional groups $\mathcal{G}_{\text{inters}} = \{G_{a_1b_1}, G_{a_1b_2}, ..., G_{a_nb_m}\}$, where each $G_{a_ib_j}$ represents individuals with a specific combination of attribute values from $A$ and $B$. This approach acknowledges the complex interplay of multiple attributes in shaping individuals' experiences and potential biases in recommendations.

\end{enumerate}

Evaluating fairness across these dimensions is crucial for ensuring that RecLLMs deliver fair and unbiased outcomes, addressing both the simplistic views of independent attributes and the complex realities of intersectional identities.

\subsection{Evaluation Method}
\label{subsec:eval_method}

This section describes our approach to evaluating the fairness of RecLLMs in generating (personalized) recommendations. Our evaluation method leverages natural language processing to understand user preferences and generate recommendations that are both sensitive and relevant to the user's interests.

\subsubsection{Data Format for User Instructions}
\label{subsec:data_format}

RecLLMs interpret user preferences expressed in natural language, enabling a personalized recommendation process in a zero-shot setting. Following the methodology similar to that described in \citet{zhang2023chatgpt} we employ and improve a \textit{structured template} designed to capture individual preferences alongside relevant sensitive attributes. This structure ensures a nuanced understanding of the user's needs. The templates are structured as follows:

\begin{enumerate}
\item \textbf{Sensitive Demographic Information.} This optional statement, tested within (\(\mathcal{R}_{m}\)) in our framework, identifies user-protected characteristics-such as age, gender, or cultural background—that might influence recommendations. It serves as the basis for defining and measuring unfairness. Examples of such statements include \textit{\dquotes{The user is female}} (individual) or \textit{\dquotes{The user is a Female Teen.}} (intersectional).

    \item \textbf{User Profile.} It is constructed based on the incorporation of two modules: \textit{the passion profile} and the \textit{item consumption profile}.
    \begin{enumerate}
        \item \textit{Passion Profile.} Considering the token length limitation for movie inclusions within ChatGPT, we design a module named \dquotes{passion profile generator} whose role is to create a narrative profile encapsulating the user's interests, often derived from their consumption history. An example of a passion profile statement taken from a random user in our dataset is: \dquotes{\textit{The user mostly likes the genres (Drama|Sci-Fi, Drama, Comedy|Romance) in the years (1951 to 1997)}}.
        \item \textit{Actual Consumption Profile.} This part provides more detailed context by detailing the genres or types of items (e.g., movies) preferred/consumed by the user, incorporating their genre and year. For instance, \dquotes{\textit{Based on the user’s preferences for the movies 'Chariots of Fire' (1981) (Genres: Drama, Rating: 5/5), 'Sabrina' (1954) (Genres: Comedy|Romance, Rating: 5/5) ...}}. It should be noted that a major aspect of our contribution includes the design and implementation of various qualitative profile construction strategies, which are tested for this section. These strategies are discussed in detail in Section \ref{subsec:Item_cons}.
    \end{enumerate}

    \item \textbf{Actual Demand Statement for Recommendations.} This specifies the user's request, often quantified by the number of recommendations sought (denoted as \(K\)), e.g., \dquotes{\textit{... Please suggest a list of 10 movie titles that the user will enjoy.}}
\end{enumerate}

\vspace{2mm}

\noindent \textbf{Example Templates for Recommendation Requests.} In light of the prompt structure detailed earlier, the following scenarios exemplify how to leverage user profiles for personalized recommendations in RecLLMs. These templates demonstrate applications both with and without sensitive attribute considerations, designated as ($\mathcal{R}_{m}^a$) and ($\mathcal{R}_{m}$) scenarios, respectively.

\begin{enumerate}
  \item  \textit{Basic Instruction Template.} \dquotes{Passion Profile + Recommendation Request}: In this basic scenario, the RecLLM exclusively leverages the user's passion profile to formulate a recommendation request.
    \begin{tcolorbox}[colback=blue!5!white,colframe=blue!75!black,title=Basic Neutral Instruction Template ($\mathcal{R}_{m}$)]
\dquotes{\textit{The user mostly likes the genres (Drama|Sci-Fi, Drama, Comedy|Romance) in the years (1951 to 1997). Please suggest a list of 10 movie titles that the user will enjoy.}}
\end{tcolorbox}

when incorporating sensitive information, the recommendation process adapts to account for this additional context,
    \begin{tcolorbox}[colback=blue!5!white,colframe=blue!75!black,title=Basic Sensitive Instruction Template ($\mathcal{R}_{m}^a$)]
\dquotes{\textit{The user is  \textcolor{red}{\textbf{Female}}. The user mostly likes the genres (Drama|Sci-Fi, Drama, Comedy|Romance) in the years (1951 to 1997). Please suggest a list of 10 movie titles that the user will enjoy.}}
\end{tcolorbox}

\item \textit{Detailed Instruction Template.} \dquotes{Passion Profile + Items Consumption Profile + Recommendation Request}: This scenario combines the user's passion profile with their item consumption history to design a more detailed recommendation request.
\begin{tcolorbox}[colback=green!5!white,colframe=green!75!black,title=Detailed Neutral Instruction Template ($\mathcal{R}_{m}$)]
\dquotes{\textit{The user prefers genres such as Drama, Sci-Fi, Drama, Comedy, and Romance, from the years (1951 to 1997). Considering the user's enjoyment of movies like 'Chariots of Fire' (1981, Drama, Rating: 5/5), 'Sabrina' (1954, Comedy|Romance, Rating: 5/5), and 'E.T. the Extra-Terrestrial' (1982, Children's|Drama|Fantasy|Sci-Fi, Rating: 5/5), recommend 10 movies that the user will enjoy.}}
\end{tcolorbox}

similarly, as with the previous scenarios, this approach is also tested with the inclusion of sensitive information, designating it as ($\mathcal{R}_{m}^a$).

\begin{tcolorbox}[colback=green!5!white,colframe=green!75!black,title=Sensitive Detailed Instruction Template ($\mathcal{R}_{m}^a$)]
\dquotes{\textit{The user, identified as \textcolor{red}{\textbf{Female}}, has preferences for genres such as Drama, Sci-Fi, Comedy, and Romance, notably from the years (1951 to 1997). Taking into account the user's sensitive attribute (Female) and their fondness for films like 'Chariots of Fire' (1981, Drama), 'Sabrina' (1954, Comedy|Romance), and others, recommend 10 movies that align with the user's taste.}}
\end{tcolorbox}

\end{enumerate}

In our study, we evaluated the scenarios labeled as \textit{detailed neutral instruction} and \textit{sensitive detailed instruction}, which correspond to \(\mathcal{R}_{m}\) and \(\mathcal{R}_{m}^a\), respectively. It is important to highlight that we developed a suite of item profile sampling strategies to select a manageable and meaningful representation of user interest, thereby addressing the constraints presented by ChatGPT, as elaborated in Section \ref{subsec:Item_cons}.

Moreover, to consider a user characterized by an intersectional identity (e.g., a young adult female) with a passion for action and drama movies from the years 2000 to 2020. The prompt generated might be:
\begin{quote}
\dquotes{The user is \textcolor{red}{\textbf{Young Adult Female}}. She mostly likes the genres (Drama|Sci-Fi, Drama, Comedy|Romance) in the years (1951 to 1997). Considering the user's sensitive \textcolor{red}{\textbf{attribute (Young Adult Female)}} and preferences for the movies chariots of fire (1981) (Genres: Drama, Rating: 5/5), sabrina (1954) (Genres: Comedy|Romance, Rating: 5/5), ... recommend 10 movies that align with the user's taste.}
\end{quote}

This approach would enable us to audit whether recommendations change by merely incorporating sensitive characteristics, which, in this context, could be interpreted as either personalization or unfairness, depending on if and how they differ with respect to other groups. 

\subsubsection{Sampling Strategies for Item Profile Construction}\label{subsec:Item_cons}

The consumption history of an individual user might encompass over 160 movies in the \texttt{ML-1M} dataset (see Table~\ref{tbl:datasets_stats_extended}), making the inclusion of their entire viewing history, including titles and genres, impractical (and redundant) frequently surpassing the model's token limit. This situation highlights a main challenge in design of prompt-based RecLLMs: \textit{how to select a representative subset of movies to form a concise yet impactful user profile?}. We have designed a suite of item profile sampling strategies: \textit{random}, \textit{top-rated}, \textit{recent}. These strategies are specifically engineered to efficiently extract user preferences from users' extensive consumption data. The main idea behind this step comes from the practical constraints imposed by the token limitations of large language models (LLMs) such as ChatGPT.  The proposed sampling strategies are meticulously designed to sift through user consumption history, ensuring that personalization remains both personalized and achievable within the confines of technical limitations.

\begin{enumerate}
    \item \textbf{Random Sampling:}
    This strategy offers a straightforward solution to bypass the token limit issue by randomly selecting movies from a user's history. This method selects a diverse yet unpredictable representation of user preferences. \vspace{2mm}
    
   \textit{Example:} If a user has watched over 160 movies, the random sampling might select \dquotes{Inception} (2010, Sci-Fi), \dquotes{The Godfather} (1972, Drama), and \dquotes{Finding Nemo} (2003, Animation), providing a broad glimpse into varied interests.  \vspace{1.0mm}

  \item  \textbf{Top-Rated Sampling:}
This strategy prioritizes movies that the user has highly rated, under the assumption that these selections best reflect their preferences. This approach efficiently utilizes limited tokens to capture high-satisfaction items. \vspace{2mm}
   
      \textit{Example:} For the same user, the top-rated sampling could highlight \dquotes{Schindler's List} (1993, Drama, Rating: 5/5) and \dquotes{The Shawshank Redemption} (1994, Drama, Rating: 5/5), focusing on movies that are favored by the user. \vspace{1.0mm}

  \item  \textbf{Recent Sampling:}
   Adding a temporal dimension, this strategy selects movies based on their recent interaction timestamps. It assumes that the most recently rated or watched movies are more indicative of current interests, making the recommendations timely and relevant. This approach is particularly useful for capturing evolving tastes and offering up-to-date suggestions.

\end{enumerate}

By incorporating these strategies, we aim to explore and test various scenarios that involve the issue of selecting the most representative movies from a potentially voluminous history. For the sake of our experiments and given the extensiveness of these analyses, we focused our attention on creating profiles that include a fixed number of items, specifically
$N_{prof}=10$, for profile inclusion. This standardization across different sampling strategies allows for a controlled comparison, ensuring that the influence of each strategy on the recommendation quality can be accurately assessed. 

\subsubsection{Evaluation Procedure}
\label{subsec:eval_proc}

The evaluation of the CFairLLM framework is designed to  assess the fairness of recommendations generated by RecLLMs with respect to \underline{both} independent and intersectional groups. Our procedure comprises several steps, aimed at examining how well the system aligns with our fairness objectives through the lenses of neutral and sensitive attribute-impacted rankings. The steps are as follows:

\begin{enumerate}
    \item \textbf{Collect Neutral and Sensitive Recommendations}: For each user instruction set \(I_m\), generate two distinct sets of recommendations,  where $m$ is the index of instruction. 
    \begin{itemize}
        \item \(\mathcal{R}_{m}\): Obtain the top-$K$ recommendations from the neutral model, which does not consider sensitive attributes.
        \item \(\mathcal{R}_{m}^{a}~~\text{\&}~~\mathcal{R}_{m}^{ab}\): Modify \(I_m\) to include sensitive attributes, forming sensitive instructions for both independent (\(\{I_{m}^{a}\}\)) and intersectional groups (\(\{I_{m}^{ab}\}\)). Gather the top-$K$ recommendations (\(\mathcal{R}_{s}^{a}\) and \(\mathcal{R}_{s}^{ab}\)) for each.\vspace{1.5mm}
    \end{itemize}

    \item \textbf{Evaluate Similarity of Information Items (\(\mathcal{B}_{item}\))}: Measure the consistency in the recommended items between the neutral (\(\mathcal{R}_{m}\)) and sensitive attribute-influenced (\(\mathcal{R}_{m}^{a}\)) ranking lists. This step involves calculating similarity metrics (e.g., Jaccard similarity, PRAG) to identify any significant disparities, indicating potential bias.

    \item \textbf{Assess True Preference Alignment (\(\mathcal{B}_{pref}\))}: Compare the recommendations from both \(\mathcal{R}_{m}\) and \(\mathcal{R}_{m}^a\) (or \(\mathcal{R}_{m}^{ab}\))  against the user's genuine preference profile. This step is crucial for ensuring that the recommendations reflect the users' actual interests and preferences, irrespective of the inclusion of sensitive attributes.

\end{enumerate}

\subsubsection{Evaluation Metrics}
\label{subsec:eval_metrics}

The evaluation of fairness in recommendations requires a diverse set of metrics that reflect various aspects of the recommendation process. These metrics are categorized into two main parts: item similarity, true preference alignment. and genre consistency/calibration.

\paragraph{Item Similarity}
Metrics under this category assess the consistency of recommended items between the neutral ranker and the sensitive ranker, without considering the ground truth preference.

\begin{itemize}
    \item \textbf{Jaccard Similarity at K (JS@K)}: This metric is calculated as
    \begin{equation}
        JS@K = \frac{1}{M} \sum_{m=1}^{M} \frac{| \mathcal{R}_{m}^{a} \cap \mathcal{R}_{m} |}{| \mathcal{R}_{m}^{a} \cup \mathcal{R}_{m} |},
    \end{equation}
    where \(\mathcal{R}_{m}^{a}\) and \(\mathcal{R}_{m}\) are the sets of top-K recommendations for the sensitive and neutral instructions, respectively, and \(S\) is the number of instructions. Fairness is denoted as \(\Delta_{\text{JS}}\), where higher values of \(JS@K\) indicate more fairness.
    
\textbf{ \item{PRAG* Metric}}: This similarity metric is formulated by adapting the Pairwise Ranking Accuracy Gap metric, which accounts for the relative rankings between two items. Explicitly, the similarity between the neutral and sensitive groups concerning the top-$K$ recommendations by a Large Language Model is defined as:

\begin{equation}
PRAG^*@K = \frac{1}{K(K + 1)S} \sum_{m} \sum_{\substack{v_1,v_2 \in \mathcal{R}_{m}^a \\ v_1 \neq v_2}} \left( \mathbb{I}(v_1 \in \mathcal{R}_{m}) \right) \times \left( \mathbb{I}(r_{m,v_1} < r_{s,v_2}) \right) \times \left( \mathbb{I}(r_{m,v_1}^a < r_{s,v_2}^a) \right),
\end{equation}

where $\mathbb{I}(\cdot)$ retains the meaning as defined previously, $v_1$ and $v_2$ signify two distinct recommended items in $\mathcal{R}_{m}^a$ and $r_{m,v_i}$ (or $r_{m,v_i}^a$) symbolizes the rank of $v_i$ in $\mathcal{R}_{m}$ (or $\mathcal{R}_{m}^a$ respectively). In particular, if $v_1$ is not listed in $\mathcal{R}_{m}$, then $r_{m, v_1}$ is set to $+\infty$, and similarly for $v_2$. As elucidated by the formula, a higher metric value not only demands a substantial overlap of items but also necessitates that the pairwise ranking sequence of any given item relative to another must be congruent in $\mathcal{R}_{m}$ and $\mathcal{R}_{m}^a$. This criterion enables us to gauge the concordance of pairwise rankings between the recommendation outputs for both neutral and sensitive instructions.
\end{itemize}

\subsubsection{True Preference Alignment}
\label{subsubsec:true_prf}
To enhance the evaluation of recommendation systems, we adapt the approach proposed by~\citet{zhang2023chatgpt} to more model fairness in terms of how well recommendations align with users' true preferences. We leverage test data to refine the accuracy of our evaluation metrics. Specifically, we introduce two modified variables, $\mathcal{R}{m}'$ and $\mathcal{R}{m}'^a$, to denote the items within the recommendation lists that are favored by users, obtained from their interactions in the test data. Consequently, this adjustment results in a comprehensive list of refined recommendation lists.

\begin{itemize}
    \item  $\mathcal{R}_{m}$: The set of items recommended to a user $m$ by the neutral ranker.
    \item $\mathcal{R}_{m}^a$: The set of items recommended to the same user by the sensitive ranker.
    \item  $\mathcal{R}_{m}'$: The subset of $R_m$, filtered on the basis of the test data to include only those items that align with the user's true preferences.
    \item  $\mathcal{R}_{m}'^a$: Similarly, the subset of $R_m^a$, filtered to include only items genuinely preferred by the user, as per the test data.
\end{itemize}

In our evaluation of fairness based on true preference alignment, we employ the same metrics, Jaccard similarity and the PRAG metric, as means of quantifying unfairness. Essentially, we focus our attention towards \textit{the changes in high-quality items -- favored by the target user--  between two recommendation lists} to flag the system as unfair.

\noindent \textcolor{customblue}{\paragraph{Fairness Metrics}
Similar to \citet{zhang2023chatgpt}, we propose two fairness metrics — \textit{Sensitive-to-Neutral Similarity Range (SNSR)} and \textit{Sensitive-to-Neutral Similarity Variance (SNSV)}, which quantify the unfairness level by measuring 
the divergence of \(\{\text{Sim}(a) | a \in A\}\) from different aspects.}

\begin{itemize}
    \item \textcolor{customblue}{\textbf{Sensitive-to-Neutral Similarity Range (SNSR):} This metric measures the disparity in similarity scores between the most advantaged and disadvantaged sensitive groups. Formally, for the top-$K$ recommendations, it is defined as:
    \[
    SNSR@K = \max_{a \in A} \overline{\text{Sim}(a)} - \min_{a \in A} \overline{\text{Sim}(a)},
    \]
    where \(\overline{\text{Sim}(a)}\) denotes the average similarity score for the sensitive group \(a\), and \(A\) represents the set of all possible values of the studied sensitive attribute. A higher SNSR value indicates greater unfairness due to larger disparities in similarity across groups.}

    \item \textcolor{customblue}{\textbf{Sensitive-to-Neutral Similarity Variance (SNSV):} This metric captures the variability in similarity scores across all sensitive groups by computing the variance of \(\overline{\text{Sim}(a)}\). The formula for SNSV is:
    \[
    SNSV@K = \sqrt{\frac{1}{|A|} \sum_{a \in A} \left( \overline{\text{Sim}(a)} - \frac{1}{|A|} \sum_{a' \in A} \overline{\text{Sim}(a')} \right)^2},
    \]
    where \(|A|\) denotes the total number of sensitive groups in \(A\). A higher SNSV value implies greater variability in the similarity scores, reflecting higher levels of unfairness.}

\end{itemize}

\noindent \textcolor{customblue}{Both fairness metrics aim to assess the sensitivity of recommendations to different groups, with higher values indicating greater disparities or inconsistencies.}

\subsection{Setup}
\noindent \textcolor{customblue}{\subsubsection{Data.} To evaluate the effectiveness of our recommendation systems, we utilized two widely recognized datasets from different domains: \texttt{ML-1M} (movies) and \texttt{LastFM-1K} (music). These datasets were chosen because they represent distinct domains and contain sensitive attributes, making them suitable for our analysis.}

\textcolor{customblue}{For \texttt{ML-1M}, we used users' explicit movie ratings, while for \texttt{LastFM-1K}, we started with implicit feedback in the form of user-song play counts. To make this dataset more similar to the movie domain, the implicit feedback was converted to explicit feedback on a scale of 1 to 5, following the procedure proposed in \cite{deldjoo2020dataset,lee2015escaping}. This conversion ensured consistency in prompt construction and allowed for a more direct comparison between the two datasets.} 

\textcolor{customblue}{We selected not to use the \texttt{LastFM-360K} dataset, as it is limited to artist-level interactions, whereas \texttt{LastFM-1K} provides both artist and song-level data, making it better suited for our study. For each dataset, we partitioned the data into training and testing sets using a temporal splitting strategy. Users' own ratings or interactions were used to construct their profiles in the training set, while the test set was reserved for evaluation. The statistics of the final datasets are shown in Table \ref{tbl:datasets_stats_extended}.} \vspace{2mm}



Given the complex nature of our experiments, each user in our dataset is exposed to various scenarios (instructions). These included evaluations with and without a sensitive attribute, alongside assessments employing different user sampling strategies. This multi-faceted approach required over \textit{ten distinct} instructions per target user, significantly increasing the communication and labor requirements when interfacing with OpenAI. To mitigate these challenges and manage associated costs, a strategic sampling method was adopted, where we used a subset of a representative group of 150 users from the dataset. For each  user, we repeated the specified tasks, to make a right balance between thoroughness and efficiency in our analysis.
\vspace{2mm}

\noindent \subsubsection{Sampling Strategies and Sensitive Attributes.} Our exploration of sampling strategies was comprehensive, including options such as \squotes{random}, \squotes{top-rated}, \squotes{recent}, mentioned in Section~\ref{subsec:Item_cons} In parallel, we examined various sensitive attributes to understand their impact on recommendation fairness. These attributes included \squotes{gender,} \squotes{age group,} \squotes{intersectional}, and scenarios excluding sensitive attributes altogether. \\


\noindent \subsubsection{NLP processing and databse search} In our methodology, we address the challenge of accurately searching and identifying movie titles within a large catalog of movie database (with titles, genre, year). 

After an initial text manipulation, we employ a \textit{regular expression-based} approach to extract movie titles and their respective release years from the structured text. This extraction process is designed to accommodate various formats in which movie information might be presented, thereby enhancing the flexibility and robustness of our method. The core of our search algorithm, similar to \cite{di2023evaluating}, utilizes the \squotes{difflib} library,\footnote{\url{https://docs.python.org/3/library/difflib.html}} a Python module known for its capability to perform sequence matching. \squotes{difflib} enables us to find the best match for each movie title within the database by comparing the preprocessed titles against the titles stored within the database. We apply a threshold for match similarity to ensure a high degree of accuracy in the results. \noindent \textcolor{customblue}{We performed a very similar post-processing step for music data.}

To accommodate the nuances of human language and potential discrepancies in movie title representations, our methodology includes a step for converting all titles to lowercase and stripping any leading or trailing whitespace. This normalization process ensures that the comparison between the input titles and the database is not hindered by case sensitivity or extraneous characters. The outcome of this process is a list of matched titles, each associated with its corresponding unique identifier within the database. This enables a seamless integration of the search results with further analytical or recommendation-based processes, thereby contributing to the overall objective of enhancing movie discovery and recommendation systems.

\begin{table}[!t]
\caption{\textcolor{customblue}{Statistics of the datasets used in our work.}}
\label{tbl:datasets_stats_extended}
\centering
\begin{tabular}{lcccccc}
\toprule
\textbf{Dataset} & \textbf{|U|} & \textbf{|I|} & \textbf{|R|} & \textbf{Density (\%)} & \textbf{$\frac{R}{U}$} & \textbf{$\frac{R}{I}$} \\
\midrule
\textbf{\texttt{ML-1M} (train)} & 150 & 2,537 & 18,428 & 95.16 & 122.85 & 7.26 \\
\textbf{\texttt{ML-1M} (test)}  & 150 & 1,590 & 4,023  & 98.31 & 26.82  & 2.53 \\
\textbf{\texttt{LastFM-1K} (train)}    & 149 & 21,967 & 37,534 & 98.85 & 251.91 & 1.71 \\
\textbf{\texttt{LastFM-1K} (test)}     & 150 & 7,308  & 9,460  & 99.14 & 63.07  & 1.29 \\
\bottomrule
\multicolumn{7}{p{0.65\linewidth}}{\textcolor{customblue}{\textbf{Note:} We used a temporal splitting strategy, so that train ratings/interactions are only used for profile construction.}}
\end{tabular}
\end{table}


\section{Results and Discussion}
\noindent \textcolor{customblue}{\textbf{Experimental Research Questions.} Throughout this section, our objective is to answer the following set of experimental research questions.}

\begin{itemize}
    \item \textcolor{customblue}{\textbf{RQ1:}  How does true preference alignment compare to similarity-based alignment in measuring fairness within RecLLMs? }
    \vspace{0.35mm}
 \item \textcolor{customblue}{\textbf{RQ2:} How do user profile sampling strategies impact fairness and accuracy in RecLLMs?}
 \item \textcolor{customblue}{\textbf{RQ3:} How does the variability in consumer fairness measures differ across multiple RecLLM \textit{models} and \textit{datasets}, and to what extent does the introduction of sensitive attributes—especially intersectional attributes—amplify observed disparities in fairness?}
 \item \textcolor{customblue}{\textbf{RQ4:}  How does increasing the scope of sampling strategies ($N$ number of selected movies) impact the computed similarities and fairness?}
\end{itemize}

\textcolor{customblue}{\subsection{Answer to RQ1. How does true preference alignment compare to similarity-based alignment in measuring fairness within RecLLMs?}}

\begin{figure}[ht!]
    \centering

    \begin{subfigure}[b]{\textwidth}
        \centering
        \includegraphics[width=0.8\textwidth]{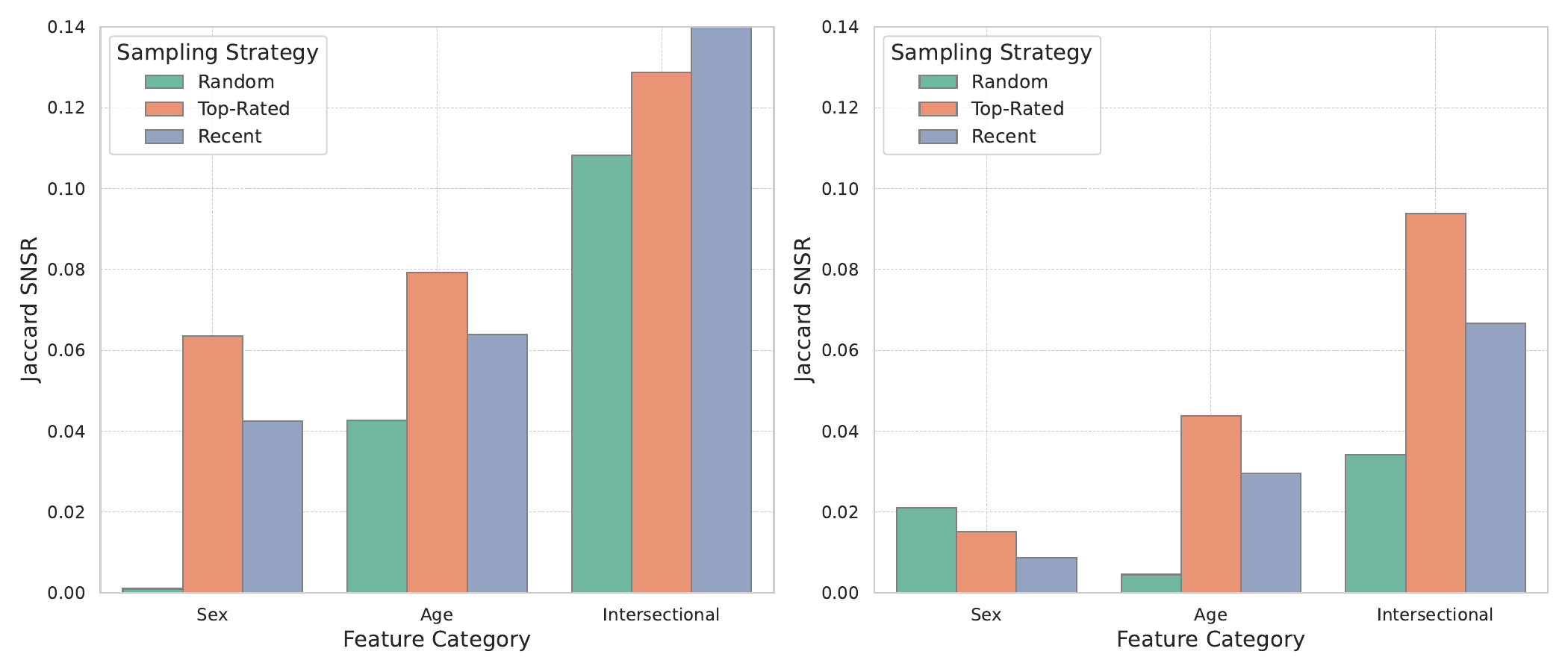}
        \caption{Jaccard SNSR, Item-Level Similarity (Left) and True Preference Alignment (Right)}
        \label{fig:item_similarity2}
    \end{subfigure}
    \vspace{0.05cm} 

    \begin{subfigure}[b]{\textwidth}
        \centering
        \includegraphics[width=0.8\textwidth]{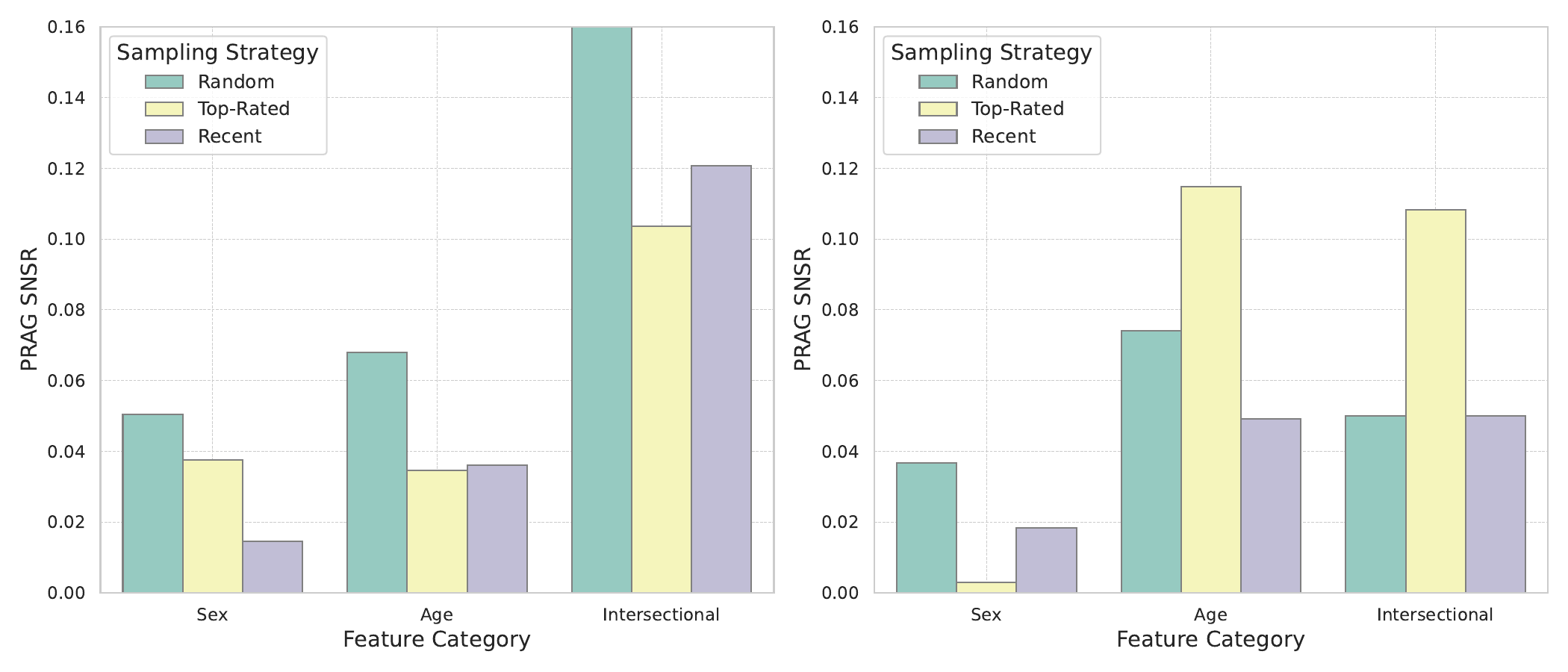}
        \caption{PRAG SNSR,  Item-Level Similarity (Left), True Preference Alignment (Right)}
        \label{fig:true_preference_alignment2}
    \end{subfigure}

    \caption{Fairness and Accuracy Metrics Across Sampling Strategies on \texttt{ML-1M} dataset.}
    \label{fig:fairness_accuracy2}
\end{figure}

\textcolor{customblue}{Results for this RQ are summarized in Table \ref{tab:sim_align5}, and Figure \ref{fig:fairness_accuracy2}. The comparison between true preference alignment (\(\beta_{pref}\)) and similarity-based alignment (\(\beta_{item}\)) reveals distinct trade-offs in understanding and evaluating fairness within RecLLMs. Overall, true preference alignment consistently results in lower similarity scores across all sampling strategies and sensitive attribute groups, indicating a divergence between the two recommendation approaches. For instance, under \textit{random sampling}, the Jaccard similarity for the Sex category is \(0.0313\) for \(\beta_{pref}\) compared to \(0.1680\) for \(\beta_{item}\). Similarly, in the Age category, true preference alignment yields similarity scores of \(0.0226\) (Teen), \(0.0199\) (Young), and \(0.0181\) (Adult), markedly lower than their \(\beta_{item}\) counterparts of \(0.1669\), \(0.1847\), and \(0.1421\) respectively. This reduction underscores that true preference alignment tailors recommendations more closely to individual user interests, potentially enhancing personalized relevance at the expense of broader item similarity.}

\textcolor{customblue}{Despite the lower similarity scores, true preference alignment demonstrates superior fairness as evidenced by the SNSR and SNSV metrics. For example, under \textit{random sampling} in the Sex category, \(\beta_{pref}\) exhibits an SNSR of \(0.0210\) and SNSV of \(0.0105\), compared to \(\beta_{item}\)'s SNSR of \(0.0010\) and SNSV of \(0.0005\). Although higher SNSR and SNSV values generally indicate greater fairness discrepancies, the overall context suggests that true preference alignment mitigates unfair biases by aligning recommendations more closely with genuine user preferences, thereby promoting equitable treatment across different user groups. For instance, in all the tested scenarios, the SNSR and SNSV values of \(\beta_{pref}\) are consistently 3 to 10 times lower than those of \(\beta_{item}\). For example, considering SNSV across the Sex, Age, and Intersectional categories, the respective values are (0.0076 vs. 0.0318), (0.0179 vs. 0.0370), and (0.0305 vs. 0.0501). This highlights that, when fairness is assessed based on underlying norms and the sensitive attributes at stake, true preference alignment demonstrates relatively lower levels of unfairness, a discussion overseen in the findings of \citet{zhang2023chatgpt}.}

\textcolor{customblue}{A similar trend is evident in Figure \ref{fig:fairness_accuracy2} (compare the left and right panels). Switching to true preference alignment reduces (in some case) the SNSR by approximately half (both for Jaccard and PRAG). This demonstrates that emphasizing whether recommendations truly match user preferences significantly diminishes disparities across different age groups.}

\begin{tcolorbox}[colback=gray!25!white,colframe=gray!75!black]
\textbf{Summary of Answer to RQ1.} 

\textcolor{customblue}{In summary, true preference alignment in RecLLMs offers both more personalized and fairer recommendations by closely aligning with individual user preferences, albeit at the expense of lower overall item similarity. As hypothesized earlier in the paper, the perception of fairness and unfairness is significantly influenced by the norms and values considered, as well as how these norms define and measure unfairness \cite{deldjoo2024normative}.}

\end{tcolorbox}

\begin{table}[!t]
\caption{Recommendation alignment between  ($\mathcal{R}_{m}$, ~$\mathcal{R}_{m}^a$) based on Item Similarity \(\beta_{item}\) and True Preference Alignment \(\beta_{pref}\). Detail results can be found in Appendix, Table \ref{tab:sim_align} and \ref{tab:pref_align}.}
\small
\label{tab:sim_align5}
\centering
\renewcommand{\arraystretch}{1.50} 
\resizebox{1.02\textwidth}{!}{%
\begin{tabular}{lcccccccccccccccccccc}
\toprule
\makecell{\small{\textbf{Sim.}}}  & \makecell{\small{\textbf{Profile}} \\ \small{\textbf{Sampling}}} & \multicolumn{4}{c}{\textbf{\textcolor{red}{\large{\textbf{Sex}}}}} & & \multicolumn{5}{c}{\textbf{\textcolor{red}{\large{\textbf{Age}}}}} & & \multicolumn{7}{c}{\textcolor{red}{\large{\textbf{Intersectional (Sex \& Age)}}}} \\
\cline{3-6} \cline{8-12} \cline{14-21}
& & $S_{mn}$ & $S_{fn}$ &\textbf{SNSR}~$\downarrow$ &\textbf{SNSV}~$\downarrow$ & & $S_{tn}$ & $S_{yn}$ & $S_{an}$ &\textbf{SNSR}~$\downarrow$ &\textbf{SNSV}~$\downarrow$ & & $S_{mtn}$ & $S_{myn}$ & $S_{man}$ & $S_{ftn}$ & $S_{fyn}$ & $S_{fan}$ &\textbf{SNSR}~$\downarrow$ & \textbf{SNSV}~$\downarrow$ \\
\toprule
\multicolumn{21}{c}{\large{Jaccard}} \\
\toprule    
\multirow{3}{*}{\textcolor{red}{\textbf{$(\beta_{item})$}}} & \textbf{\small{random}} & \colorbox{purple!30}{0.1680} & \colorbox{white}{0.1670} & \colorbox{purple!30}{0.0010} & \colorbox{purple!30}{0.0005} & & \colorbox{purple!30}{0.1669} & \colorbox{purple!30}{0.1847} & \colorbox{purple!30}{0.1421} & 0.0426 & 0.0175 & & 0.2047 & 0.1532 & 0.1630 & 0.1794 & 0.1256 & 0.0965 & 0.1082 & 0.0351 \\
& \textbf{\small{top-rated}} & \colorbox{purple!30}{0.6760} & \colorbox{white}{0.6125} & \colorbox{white}{0.0635} & \colorbox{purple!30}{0.0318} & & \colorbox{white}{0.5773} & \colorbox{white}{0.6548} & \colorbox{white}{0.6565} & 0.0793 & \colorbox{purple!30}{0.0370} & & 0.4734 & 0.5905 & 0.6021 & 0.5355 & 0.5297 & 0.4743 & 0.1288 & \colorbox{purple!30}{0.0501} \\
& \textbf{\small{recent}} & \colorbox{white}{0.6344} & \colorbox{white}{0.5920} & \colorbox{white}{0.0424} & 0.0212 & & 0.5918 & 0.6556 & 0.6157 & 0.0639 & 0.0263 & & 0.5499 & 0.6004 & 0.6065 & 0.4623 & 0.5675 & 0.4741 & 0.1442 & 0.0566 \\
\midrule
\multirow{3}{*}{\textcolor{red}{\textbf{$(\beta_{pref})$}}} & \textbf{\small{random}} & \colorbox{purple!30}{0.0313} & \colorbox{white}{0.0103} & \colorbox{purple!30}{0.0210} & \colorbox{purple!30}{0.0105} & & \colorbox{purple!30}{0.0226} & \colorbox{purple!30}{0.0199} & \colorbox{purple!30}{0.0181} & {0.0045} & {0.0019} & & 0.0342 & 0.0222 & 0.0325 & 0.0317 & 0.0159 & 0.0000 & 0.0342 & 0.0120 \\
& \textbf{\small{top-rated}} & \colorbox{purple!30}{0.0681} & \colorbox{white}{0.0529} & 0.0152 & \colorbox{purple!30}{0.0076} & & 0.0919 & 0.0481 & 0.0688 & 0.0437 & \colorbox{purple!30}{0.0179} & & 0.0919 & 0.0497 & 0.0744 & 0.1063 & 0.0567 & 0.0125 & 0.0938 & \colorbox{purple!30}{0.0305} \\
& \textbf{\small{recent}} & 0.0461 & 0.0375 & \colorbox{white}{0.0086} & 0.0043 & & 0.0658 & 0.0445 & 0.0362 & 0.0296 & 0.0125 & & 0.0825 & 0.0350 & 0.0358 & 0.0159 & 0.0398 & 0.0250 & 0.0666 & 0.0210 \\
\bottomrule
\end{tabular}}
\end{table}

\textcolor{customblue}{\subsection{Answer to RQ2. How do user profile sampling strategies and their scope impact fairness and accuracy in RecLLMs?}}

\textcolor{customblue}{When we examine fairness using SNSR (where lower is better) across different sampling strategies (e.g., \textit{random}, \textit{top-rated}, \textit{recent}) and metrics (e.g., Jaccard vs. PRAG), no single strategy consistently outperforms the others in all scenarios, as shown in Figure \ref{fig:fairness_accuracy2}. For instance, under item-level similarity (Jaccard), a \textit{random} sampling strategy often presents the lowest SNSR values, suggesting it excels in surface-level fairness. However, in ranking-based scenarios (PRAG), \textit{random} performs the worst across the three feature categories, indicating a lack of precision when ranking is considered. In these ranking scenarios, \textit{top-rated} and \textit{recent} strategies demonstrate stronger performance, offering better alignment across sensitive groups.}

\textcolor{customblue}{When we focus on true preference alignment (right side of the plots), the outcomes differ. For Jaccard, \textit{random} again performs best, showing its ability to increase hits and capture user preferences broadly. However, in ranking-based evaluations (PRAG), \textit{random} typically falls behind \textit{recent}, which achieves lower SNSR values, reflecting its strength in scenarios where nuanced understanding of user profiles is essential for fair ranking.}

\textcolor{customblue}{Thus, we observe that while \textit{random} sampling  increases the number of hits and excels in surface-level item similarity (Jaccard), it has the worst impact on ranking-based fairness (PRAG), where a more careful understanding of user profiles is necessary. This arguable may highlight the importance of selecting sampling strategies tailored to the specific fairness objectives being prioritized.} \vspace{5mm}

\begin{tcolorbox}[colback=gray!25!white,colframe=gray!75!black]
\textbf{Summary of Answer to RQ2.} 

\textcolor{customblue}{In summary, each sampling strategy exhibits different strengths depending on how fairness is defined and measured. \textit{Top-rated} and \textit{recent} strategies are better suited for ranking-based scenarios (PRAG), while a simple \textit{random} strategy performs better for hits (Jaccard). Ultimately, the choice of strategy and evaluation lens (item-level vs. true preference, Jaccard vs. PRAG) determines which approach appears \dquotes{best,} and no single method universally outperforms the others in all fairness scenarios. These findings suggest that much of the fairness discussion in RecLLMs may depend on bridging the semantic gap between systems' understanding of user tastes and interest.}

\end{tcolorbox}

\begin{figure}[ht!]
    \centering

    \begin{subfigure}[b]{\textwidth}
        \centering
        \includegraphics[width=0.850\textwidth]{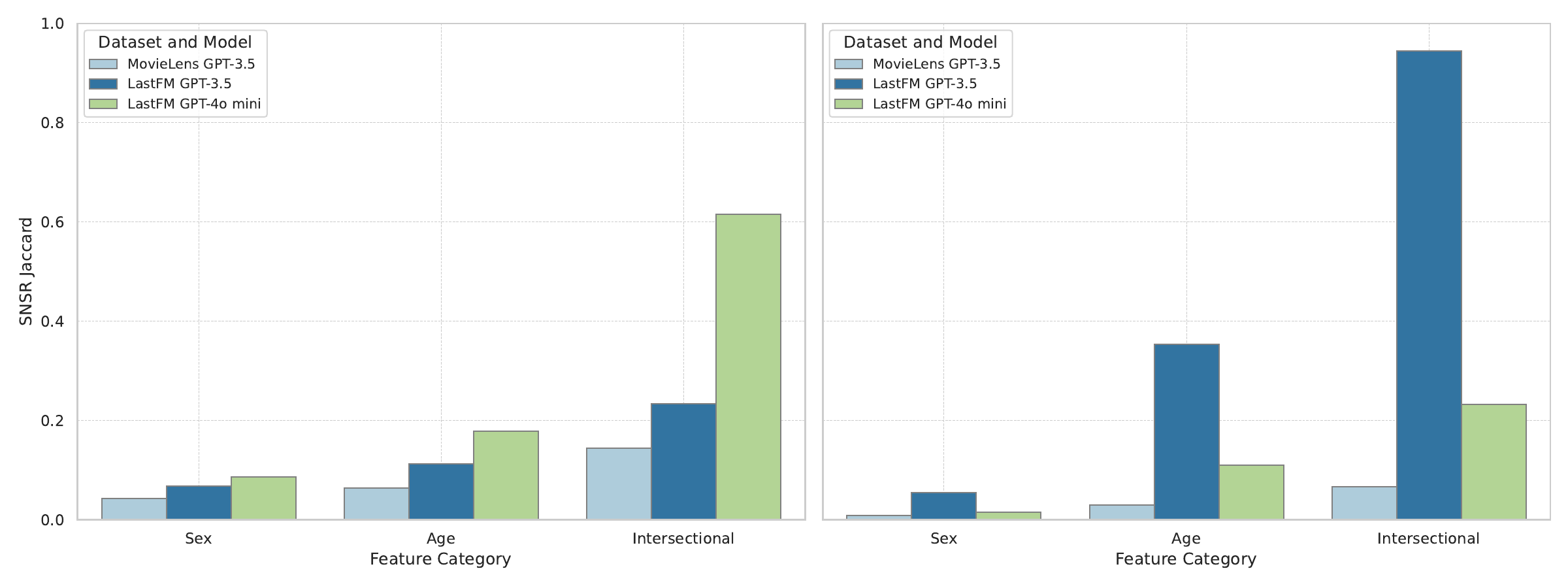}
        \caption{Jaccard SNSR, Item-Level Similarity (Left) and True Preference Alignment (Right)}
        \label{fig:item_similarity3}
    \end{subfigure}
    \vspace{0.05cm} 

    \begin{subfigure}[b]{\textwidth}
        \centering
        \includegraphics[width=0.850\textwidth]{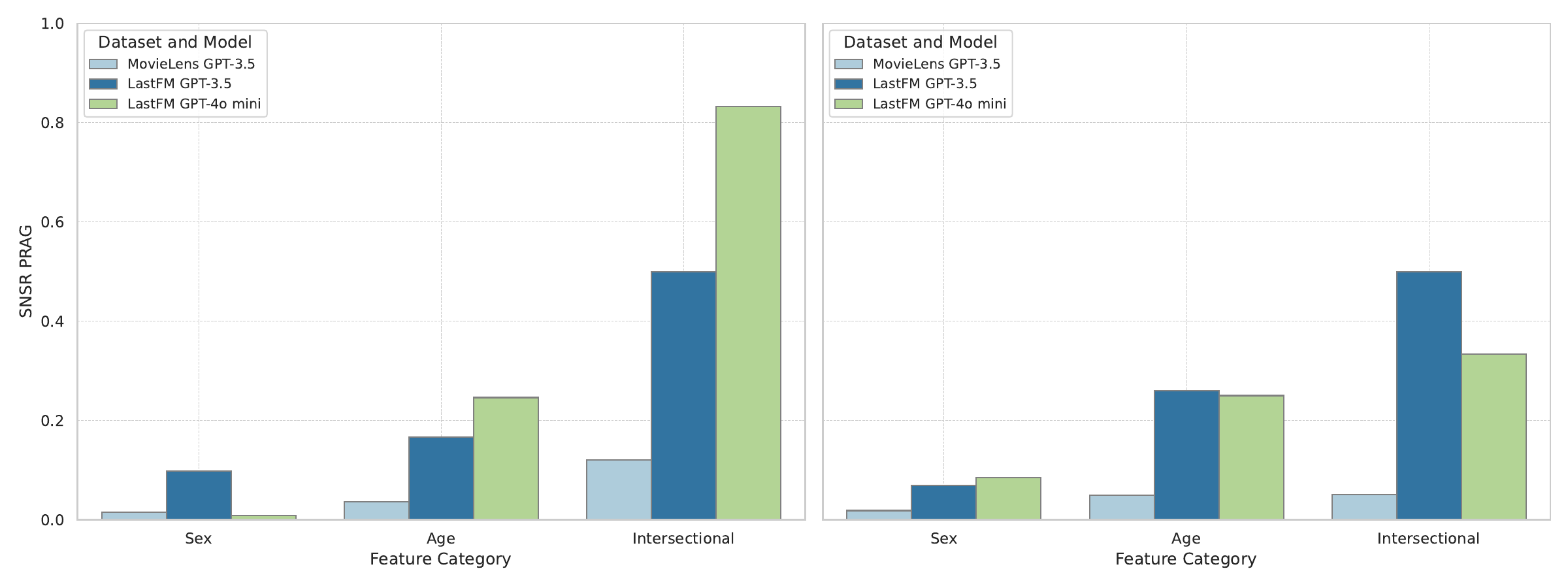}
        \caption{PRAG SNSR,  Item-Level Similarity (Left), True Preference Alignment (Right)}
        \label{fig:true_preference_alignment3}
    \end{subfigure}

    \caption{Fairness and Accuracy Metrics Across Models and Datasets.}
    \label{fig:fairness_accuracy3}
\end{figure}

 \textcolor{customblue}{\subsection{Answer to RQ3. Variability in consumer fairness across \textit{models} and \textit{datasets}, and overall what extent does the introduction of \textit{sensitive attributes}—amplify observed disparities in fairness?}}

\noindent \textcolor{customblue}{Extending our exploration beyond the earlier questions, RQ3 focuses on how fairness varies across different RecLLM \textit{models} and \textit{datasets}. Previous examinations often focused on a single model or domain, but here we aim to  compare multiple configurations—for example, contrasting GPT-3.5 with GPT-4.0 mini recommendations—and apply them to different datasets, such as MovieLens (movies) and LastFM (music). By doing so, RQ3 probes whether particular model architectures or data sources inherently lead to larger fairness gaps when sensitive attributes (e.g., age or sex) are introduced. This question is especially pertinent since models trained on distinct text corpora or using varied prompt-engineering strategies may respond differently to sensitive demographic cues, either amplifying or attenuating underlying biases. Results are shown in Figure \ref{fig:fairness_accuracy3}. } 

\textcolor{customblue}{Two key observations emerge from comparing the results:}

\begin{enumerate}

    \item \textcolor{customblue}{\textbf{Dataset Differences.} Across all tested dimensions, SNSR metrics (both for Jaccard and PRAG) generally indicate lower unfairness levels on the MovieLens dataset than on LastFM, regardless of the model type or the sensitive feature considered. One possible interpretation is that MovieLens—focusing on a well-structured and broadly familiar domain such as movies—allows LLMs to more consistently maintain fair recommendations when sensitive attributes are revealed. The system may more easily base user tastes in well-defined genres, directors, or production years, reducing the risk that sensitive attributes (e.g., age group or sex) will skew the final output. In contrast, music recommendations in LastFM may be more challenging to stabilize. Taste in music can be more individualized and harder to categorize, potentially causing larger shifts in recommended items once the model factors in sensitive demographics. In this scenario, the inclusion of sensitive attributes might push the LLM to rely on stereotypical assumptions or less robust associations, leading to greater unfairness.}

    \textcolor{customblue}{For example, consider a female user in the MovieLens dataset who previously liked a variety of action and drama films. When we reveal her gender, the GPT-based recommender still suggests a balanced set of high-quality, relevant films—only slightly different from those recommended in a gender-neutral scenario. By comparison, on LastFM, revealing that a user is a \dquotes{young adult female} might cause the model to swing more heavily toward certain music genres \textit{stereotypically} associated with that demographic, causing more pronounced shifts in the top recommended artists. }

\item \textcolor{customblue}{\textbf{Model Configuration and True Preference Alignment.} Another key finding is that fairness outcomes differ when we shift from surface-level similarity metrics (which merely check if recommended items match those originally suggested) to true preference alignment metrics (which check if recommendations genuinely align with the user’s known interests). Under this more stringent evaluation, GPT-4.0 mini generally exhibits lower unfairness than GPT-3.5, even though at a surface level the reverse seems to be true. For instance, GPT-3.5 might maintain a more consistent set of items between neutral and sensitive conditions at first glance, but closer inspection reveals that it  does so by suggesting less truly preferred items. GPT-4.0 mini, on the other hand, might appear to vary recommendations more initially, but it ultimately provides suggestions that better match the user actual tastes when sensitive attributes are considered.}

\textcolor{customblue}{The interpretation of the above can be as follows. Suppose a user who enjoys classic rock and indie bands receives recommendations from GPT-3.5 that appear stable whether we mention their age group or not. At first glance, this stability seems fair. However, looking at the user's actual listening habits in the test data, we might realize that GPT-3.5’s sensitive-attribute-influenced list includes fewer bands the user actually likes. GPT-4.0 mini’s recommendations, while more noticeably changing once the age attribute is introduced, may result in being closer to the user's real tastes—i.e., fewer \dquotes{filler} items and more genuinely preferred bands. Thus, from a \dquotes{true preference} fairness perspective, GPT-4.0 mini proves more equitable.}

\end{enumerate}

\vspace{5mm}
\noindent \textcolor{customblue}{Finally, we Now we aim to focus our attention on another key questions, which Sensitive Attribute Produces More Unfairness?}

\textcolor{customblue}{When we look at the charts, we typically find that intersectional attributes (e.g., combining age and sex) result in the greatest disparities. For instance, a user described simply as \dquotes{Female} may experience a slight shift in recommendations, but a user described as a \dquotes{Young Adult Female} might trigger a more pronounced change—significantly lower similarity or alignment with true preferences. Such intersectional cues provide the model with more demographic signals, potentially prompting stronger stereotypical assumptions. For example, a \dquotes{Young Adult Female} in the music domain might push the recommender to heavily favor trending pop artists, moving it away from niche, user-preferred indie bands. }

\begin{tcolorbox}[colback=gray!25!white,colframe=gray!75!black]
\textbf{Summary of Answer to RQ3.} 

\textcolor{customblue}{In summary, RQ3 reveals that certain dataset-model combinations and particularly intersectional attributes exacerbate fairness gaps. MovieLens tends to remain more stable, while LastFM exhibits greater sensitivity to demographic signals. Moreover, while GPT-3.5 may look stable at a surface level, GPT-4.0 mini ultimately shows fairer outcomes when we consider true user preferences. Intersectional attributes, adding multiple layers of demographic identity, often produce the most pronounced unfairness, as they allow models to lean more heavily on demographic stereotypes, especially in domains with less stable grounding.}

\end{tcolorbox}

\subsection{RQ4: The impact of Scope of The Sampling Strategy and Its Interplay with Sampling Strategy Itself}

\begin{figure}[!b]
    \centering
    \includegraphics[width = 0.9\textwidth]{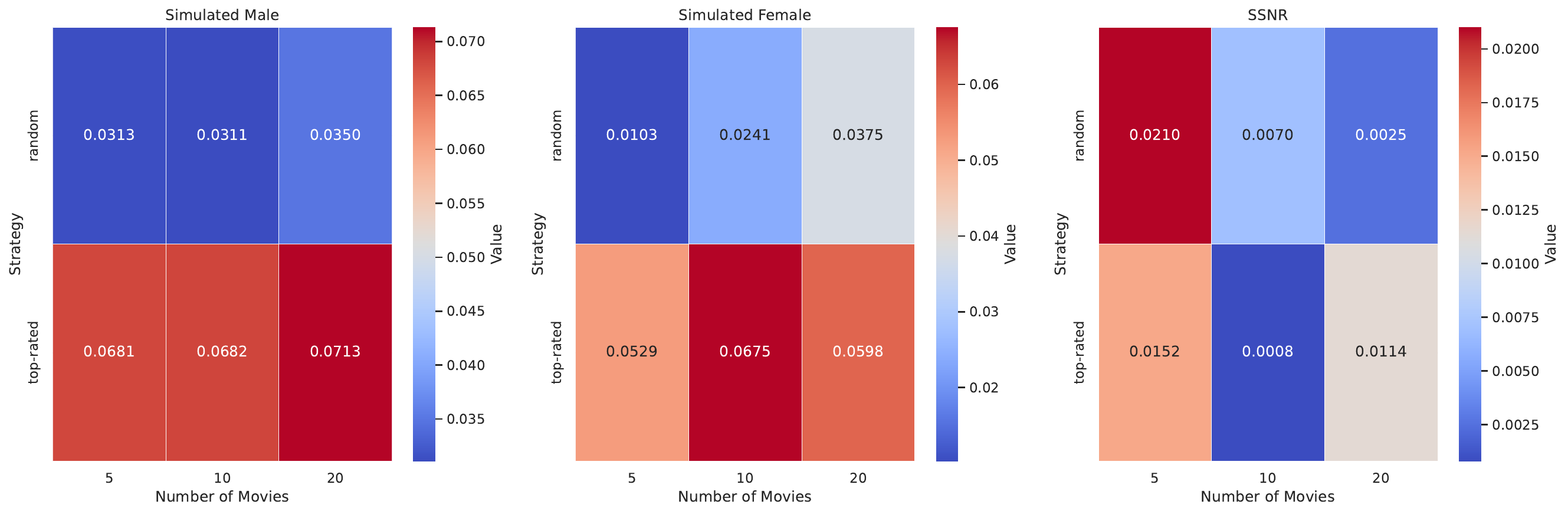}
    \caption{Heatmap Comparison of Recommendation Fairness and Similarity: The left heatmap shows the effect of increasing movie counts with random sampling, the middle heatmap depicts the outcome of using a top-rated sampling strategy, and the right heatmap presents fairness scores, highlighting the differential impact of sampling strategies on recommendation quality.}
    \label{fig:heatmaps}
\end{figure}
Increasing the number of movies in the sampling strategies has an impact on the computed similarities and fairness measures. The heatmap comparisons reveal a direct relationship between the sampling scope (\squotes{N}) and outcome metrics for both male and female profiles. While the scope (\squotes{N}) and the type of sampling strategy (top-rated vs. random) both play crucial roles, the strategy itself exhibits a more pronounced impact. Notably, the fairness values substantially decrease, and similarities increase across all cases when observing the heatmaps vertically. However, the impact of N on fairness remains slightly variable, with $N=10$ often resulting in the lowest (hence, best) fairness scores, while scores for $N=5$ and $N=15$ are slightly higher. Nonetheless, focusing on their interplay, it becomes clear that higher values of \squotes{N}, especially with the \underline{top-rated} sampling strategy, yield much better recommendation fairness scores. 

The results indicate that enhanced profile construction strategies can mitigate fairness issues arising from the introduction of sensitive attributes to RecLLMs. This phenomenon could be explained due to improved system personalization. By refining how profiles are constructed, it is possible to achieve a more nuanced understanding of user preferences, leading to recommendations that are more fair and more personalized.

\begin{tcolorbox}[colback=gray!25!white,colframe=gray!75!black]
\textbf{Answer to RQ4.} \\
The findings from RQ4 underscore the impact of both the scope of sampling strategies and the choice of sampling strategy itself on the fairness and accuracy of recommendations in RecLLMs. \underline{More detailed} profile construction strategies, particularly when increasing the scope for the user profile, and top-rated sampling (as opposed to random), impact fairness and personalization.
\end{tcolorbox}

\section{Conclusion}
\label{sec:conc}

\noindent \textcolor{customblue}{In this study, we critically examine and advance the methodologies for evaluating fairness in Large Language Model-based recommender systems (RecLLMs). Traditional approaches have predominantly focused on assessing consumer fairness by comparing recommendation lists generated with and without sensitive user attributes. However, such methods often conflate genuine personalization with biased outcomes, failing to discern whether discrepancies arise from true alignment with user preferences or from inherent biases. To address these limitations, we introduce \textbf{CFaiRLLM,} an enhanced evaluation framework that not only emphasizes \textit{true preference alignment} but also rigorously investigates \textit{intersectional fairness} by considering overlapping sensitive attributes. Additionally, CFaiRLLM incorporates diverse user profile sampling strategies—random, top-rated, and recency-focused—to address the token limitations of LLMs, aiming for a more comprehensive and realistic assessment of fairness within RecLLMs.}

\textcolor{customblue}{Our extensive experiments utilizing the \texttt{ML-1M} and \texttt{LastFM-1K} datasets reveal that true preference alignment significantly improves the personalization and fairness of recommendations compared to traditional similarity-based measures. Notably, our findings demonstrate that intersectional attributes exacerbate fairness gaps, particularly in less structured domains such as music recommendations. These insights underscore the necessity for future fairness evaluations in RecLLMs to prioritize true preference alignment, thereby fostering equitable and genuinely personalized recommendation experiences. By refining the evaluation framework and highlighting the complex interplay of intersectional identities, our work lays a foundational path for developing more ethical and user-centric recommender systems in the era of large language models.}

\textcolor{customblue}{\textbf{Key Insights from Results:}}
\begin{itemize}
   \item \textcolor{customblue}{\textbf{True Preference Alignment Enhances (our Understanding of) Fairness}: Evaluations based on true preference alignment consistently showed lower unfairness scores compared to traditional similarity-based metrics, indicating a more accurate reflection of user preferences.}
\item \textcolor{customblue}{\textbf{Intersectional Attributes Amplify Fairness Gaps:} Incorporating multiple sensitive attributes simultaneously led to more pronounced disparities in recommendation fairness, especially noticeable in the LastFM dataset.}
\item \textcolor{customblue}{\textbf{Sampling Strategies Influence Fairness Outcomes:} Top-rated and recency-focused sampling strategies outperformed random sampling in reducing bias and improving the alignment of recommendations with user preferences.}
\item \textcolor{customblue}{\textbf{Domain-Specific Variations:} Fairness improvements were more significant in structured domains like movie recommendations (ML-1M) compared to less structured ones like music (LastFM-1K), highlighting the influence of domain characteristics on fairness outcomes.}

\end{itemize}

\textcolor{customblue}{These findings advocate for a shift in fairness evaluation paradigms towards frameworks that prioritize genuine user preference alignment and account for the multifaceted nature of user identities, thereby promoting more equitable and personalized recommender systems. In the end, we would like to emphasize that the insights and findings of this work are specific to the audited system. Additionally, we clarify that prompt engineering is not a universal solution but rather one component of a broader framework. We hope this work offers valuable insights into the evaluation of recommender systems from a trust perspective. We plan to continue this research by exploring \textbf{red teaming} techniques for increasingly robust models , focusing on aspects such as security \citet{nazary2025poison}, privacy, hallucination, and emerging fairness scenarios.}

\subsection{Limitation and Future Directions}
\textcolor{customblue}{We acknowledge that using a commercial API, such as GPT, limits reproducibility due to its proprietary nature and ongoing updates. While our previous work \cite{deldjoo2024understanding} shows semantic consistency over short time frames, future work should explore open-access models like LLama, Bloom, etc. to promote reproducibility. Additionally, we recognize that our sampling strategies—random, top-rated, and recent—do not explicitly address the distinctiveness of highly-rated items, which could obscure individual user nuances. For instance, users with high ratings for widely popular movies like Star Wars or Titanic may not have their unique preferences fully captured. Incorporating diversity-focused sampling is an important direction for future research.}

\textcolor{customblue}{Furthermore, future work could extend this study by exploring additional sensitive attributes like ethnicity or socioeconomic status, applying the CFaiRLLM framework to diverse domains such as e-commerce or music, and conducting longitudinal studies to assess the impact of fairness on user satisfaction. Bias mitigation remains another critical avenue, including developing bias-aware datasets and integrating fairness considerations into model training. Additionally, leveraging content-based models offers potential for alignment with RecLLMs, given their reliance on the target user's own rating (just like how RecLLMs function) \cite{deldjoo2018content,de2015semantics}. Finally, investigating alternative approaches to intersectional fairness, is another interesting future direction, for understanding and addressing unfairness in more complex identity overlaps (see \citet{yang2020fairness}.)}

\section{Appendix}
\textcolor{customblue}{Here we provide detailed experimental results for movielens dataset that substantiate our research findings across previous questions. The data clearly demonstrate that user profile sampling strategies—impact recommendation fairness and alignment with true user preferences compared to random sampling. Additionally, our intersectional analysis reveals that combining multiple sensitive attributes, such as sex and age, intensifies fairness disparities, arguably emphasizing the complex challenges in mitigating bias within RecLLMs.}

\begin{table}[!h]
\caption{Recommendation alignment between  ($\mathcal{R}_{m}$, ~$\mathcal{R}_{m}^a$) based on Item Similarity \(\beta_{item}\)}
\label{tab:sim_align}
\centering
\renewcommand{\arraystretch}{1.2} 
\resizebox{1.05\textwidth}{!}{%
\begin{tabular}{lcccccccccccccccccccc}
\toprule

\makecell{\small{\textbf{Prompting}} \\ \small{\textbf{Strategy}}}&\makecell{\small{\textbf{Profile sampling}} \\ \small{\textbf{Strategy}}} & \multicolumn{4}{c}{\textbf{Sex}} & & \multicolumn{5}{c}{\textbf{Age}} & & \multicolumn{7}{c}{\textbf{Intersectional (Sex \& Age)}} \\
 \cline{3-6} \cline{8-12} \cline{14-21}
&  & $S_{mn}$ & $S_{fn}$ & $SNSR$ & $SNSV$ & & $S_{tn}$ & $S_{yn}$ & $S_{an}$ & $SNSR$ & $SNSV$& & $S_{mtn}$ & $S_{myn}$ & $S_{man}$ & $S_{ftn}$ & $S_{fyn}$ & $S_{fan}$ & $SNSR$ & $SNSV$  \\
\toprule
\multicolumn{21}{c}{\large{Jaccard}} \\
\toprule    
 \multirow{3}{*}{\textcolor{red}{\large{\textbf{Sex}}}}  &\textbf{random} &\colorbox{purple!30}{0.1680} &\colorbox{purple!30}{0.1670} &\colorbox{purple!30}{0.0010} &0.0005 & &\textcolor{gray!65}{0.2069} &\textcolor{gray!65}{0.1473} &\textcolor{gray!65}{0.1836} &\textcolor{gray!65}{0.0596} &\textcolor{gray!65}{0.0245} & &\textcolor{gray!65}{0.2098} &\textcolor{gray!65}{0.1438} &\textcolor{gray!65}{0.1850} &\textcolor{gray!65}{0.1985} &\textcolor{gray!65}{0.1552} &\textcolor{gray!65}{0.1780} &\textcolor{gray!65}{0.0661} &\textcolor{gray!65}{0.0230}  \\
  &\textbf{top-rated} &\colorbox{purple!30}{0.6760} &\colorbox{purple!30}{0.6125} &\colorbox{purple!30}{0.0635} &0.0318 & &\textcolor{gray!65}{0.6620} &\textcolor{gray!65}{0.6634} &\textcolor{gray!65}{0.6465} &\textcolor{gray!65}{0.0169} &\textcolor{gray!65}{0.0077} & &\textcolor{gray!65}{0.6581} &\textcolor{gray!65}{0.6874} &\textcolor{gray!65}{0.6669} &\textcolor{gray!65}{0.6733} &\textcolor{gray!65}{0.6107} &\textcolor{gray!65}{0.5648} &\textcolor{gray!65}{0.1226} &\textcolor{gray!65}{0.0425} \\
&\textbf{recent} &\colorbox{purple!30}{0.6344} &\colorbox{purple!30}{0.5920} &\colorbox{purple!30}{0.0424} &0.0212 & &\textcolor{gray!65}{0.6373} &\textcolor{gray!65}{0.6133} &\textcolor{gray!65}{0.6329} &\textcolor{gray!65}{0.0241} &\textcolor{gray!65}{0.0105} & &\textcolor{gray!65}{0.6534} &\textcolor{gray!65}{0.6119} &\textcolor{gray!65}{0.6627} &\textcolor{gray!65}{0.5913} &\textcolor{gray!65}{0.6163} &\textcolor{gray!65}{0.5138} &\textcolor{gray!65}{0.1489} &\textcolor{gray!65}{0.0488} \\
\midrule
\multirow{3}{*}{\textcolor{blue}{\large{\textbf{Age}}}}  &\textbf{random} &\textcolor{gray!65}{0.1696} &\textcolor{gray!65}{0.1716} &\textcolor{gray!65}{0.0020} &\textcolor{gray!65}{0.0010} & &\colorbox{purple!30}{0.1669} &\colorbox{purple!30}{0.1847} &\colorbox{purple!30}{0.1421} &0.0426 &0.0175 & &\textcolor{gray!65}{0.1670} &\textcolor{gray!65}{0.1879} &\textcolor{gray!65}{0.1387} &\textcolor{gray!65}{0.1666} &\textcolor{gray!65}{0.1778} &\textcolor{gray!65}{0.1557} &\textcolor{gray!65}{0.0492} &\textcolor{gray!65}{0.0157} \\
& \textbf{top-rated} &\textcolor{gray!65}{0.6372} &\textcolor{gray!65}{0.6524} &\textcolor{gray!65}{0.0152} &\textcolor{gray!65}{0.0076} & &\colorbox{purple!30}{0.5773} &\colorbox{purple!30}{0.6548} &\colorbox{purple!30}{0.6565} &0.0793 &0.0370 & &\textcolor{gray!65}{0.5598} &\textcolor{gray!65}{0.6391} &\textcolor{gray!65}{0.6821} &\textcolor{gray!65}{0.6272} &\textcolor{gray!65}{0.6893} &\textcolor{gray!65}{0.5543} &\textcolor{gray!65}{0.1351} &\textcolor{gray!65}{0.0530} \\
 &\textbf{recent}  &\textcolor{gray!65}{0.6162} &\textcolor{gray!65}{0.6793} &\textcolor{gray!65}{0.0631} &\textcolor{gray!65}{0.0315} & &0.5918 &0.6556 &0.6157 &0.0639 &0.0263 & &\textcolor{gray!65}{0.5702} &\textcolor{gray!65}{0.6338} &\textcolor{gray!65}{0.6138} &\textcolor{gray!65}{0.6534} &\textcolor{gray!65}{0.7036} &\textcolor{gray!65}{0.6233} &\textcolor{gray!65}{0.1334} &\textcolor{gray!65}{0.0404}\\
\midrule
\multirow{3}{*}{\textcolor{purple}{\large{\textbf{Inters.}}}}  &\textbf{random} 
&\textcolor{gray!65}{0.1655} &\textcolor{gray!65}{0.1291} &\textcolor{gray!65}{0.0364} &\textcolor{gray!65}{0.0182} & &\textcolor{gray!65}{0.1981} &\textcolor{gray!65}{0.1446} &\textcolor{gray!65}{0.1497} &\textcolor{gray!65}{0.0536} &\textcolor{gray!65}{0.0241} & &0.2047 &{0.1532} &{0.1630} &{0.1794} &{0.1256} &{0.0965} &{0.1082} &{0.0351} \\
&\textbf{top-rated}  &\textcolor{gray!65}{0.5724} &\textcolor{gray!65}{0.5199} &\textcolor{gray!65}{0.0525} &\textcolor{gray!65}{0.0263} & &\textcolor{gray!65}{0.4895} &\textcolor{gray!65}{0.5714} &\textcolor{gray!65}{0.5766} &\textcolor{gray!65}{0.0871} &\textcolor{gray!65}{0.0399} & &0.4734 &0.5905 &0.6021 &0.5355 &0.5297 &0.4743 &0.1288 &0.0501 \\
&\textbf{recent} &\textcolor{gray!65}{0.5929} &\textcolor{gray!65}{0.5313} &\textcolor{gray!65}{0.0616} &\textcolor{gray!65}{0.0308} & &\textcolor{gray!65}{0.5272} &\textcolor{gray!65}{0.5901} &\textcolor{gray!65}{0.5800} &\textcolor{gray!65}{0.0629} &\textcolor{gray!65}{0.0276} & &0.5499 &0.6004 &0.6065 &0.4623 &0.5675 &0.4741 &0.1442 &0.0566 \\

\toprule
\multicolumn{21}{c}{\large PRAG} \\
\toprule
 \multirow{3}{*}{\textcolor{red}{\large{\textbf{Sex}}}}  &\textbf{random} &0.3937 &0.3432 &\colorbox{purple!30}{0.0505} &0.0253 & &\textcolor{gray!65}{0.4125} &\textcolor{gray!65}{0.3651} &\textcolor{gray!65}{0.3887} &\textcolor{gray!65}{0.0475} &\textcolor{gray!65}{0.0194} & &\textcolor{gray!65}{0.4450} &\textcolor{gray!65}{0.3655} &\textcolor{gray!65}{0.4120} &\textcolor{gray!65}{0.3197} &\textcolor{gray!65}{0.3641} &\textcolor{gray!65}{0.2958} &\textcolor{gray!65}{0.1492} &\textcolor{gray!65}{0.0507} \\
&\textbf{top-rated} &0.8668 &0.8292 &\colorbox{purple!30}{0.0376} &0.0188 & &\textcolor{gray!65}{0.8909} &\textcolor{gray!65}{0.8586} &\textcolor{gray!65}{0.8290} &\textcolor{gray!65}{0.0619} &\textcolor{gray!65}{0.0253} & &\textcolor{gray!65}{0.8950} &\textcolor{gray!65}{0.8732} &\textcolor{gray!65}{0.8376} &\textcolor{gray!65}{0.8790} &\textcolor{gray!65}{0.8264} &\textcolor{gray!65}{0.7946} &\textcolor{gray!65}{0.1004} &\textcolor{gray!65}{0.0346} \\
&\textbf{recent} &0.8637 &0.8492 &\colorbox{purple!30}{0.0145} &0.0072 & &\textcolor{gray!65}{0.8809} &\textcolor{gray!65}{0.8565} &\textcolor{gray!65}{0.8520} &\textcolor{gray!65}{0.0289} &\textcolor{gray!65}{0.0127} & &\textcolor{gray!65}{0.8773} &\textcolor{gray!65}{0.8612} &\textcolor{gray!65}{0.8596} &\textcolor{gray!65}{0.8912} &\textcolor{gray!65}{0.8464} &\textcolor{gray!65}{0.8214} &\textcolor{gray!65}{0.0697} &\textcolor{gray!65}{0.0221} \\
 \midrule
 \multirow{3}{*}{\textcolor{blue}{\large{\textbf{Age}}}}  &\textbf{random} &\textcolor{gray!65}{0.3817} &\textcolor{gray!65}{0.4488} &\textcolor{gray!65}{0.0670} &\textcolor{gray!65}{0.0335} & &0.3679 &0.4292 &0.3613 &0.0679 &0.0306 & &\textcolor{gray!65}{0.3233} &\textcolor{gray!65}{0.4156} &\textcolor{gray!65}{0.3579} &\textcolor{gray!65}{0.4952} &\textcolor{gray!65}{0.4590} &\textcolor{gray!65}{0.3750} &\textcolor{gray!65}{0.1719} &\textcolor{gray!65}{0.0591} \\
&\textbf{top-rated}  &\textcolor{gray!65}{0.8638} &\textcolor{gray!65}{0.8681} &\textcolor{gray!65}{0.0043} &\textcolor{gray!65}{0.0021} & &0.8485 &0.8617 &0.8829 &\colorbox{purple!30}{0.0345} &0.0142 & &\textcolor{gray!65}{0.8510} &\textcolor{gray!65}{0.8505} &\textcolor{gray!65}{0.8954} &\textcolor{gray!65}{0.8414} &\textcolor{gray!65}{0.8860} &\textcolor{gray!65}{0.8330} &\textcolor{gray!65}{0.0624} &\textcolor{gray!65}{0.0230}\\
&\textbf{recent} &\textcolor{gray!65}{0.8476} &\textcolor{gray!65}{0.8858} &\textcolor{gray!65}{0.0381} &\textcolor{gray!65}{0.0191} & &0.8409 &0.8545 &0.8770 &0.0361 &0.0149 & &\textcolor{gray!65}{0.8241} &\textcolor{gray!65}{0.8429} &\textcolor{gray!65}{0.8707} &\textcolor{gray!65}{0.8888} &\textcolor{gray!65}{0.8799} &\textcolor{gray!65}{0.9023} &\textcolor{gray!65}{0.0782} &\textcolor{gray!65}{0.0268} \\
\midrule
 \multirow{3}{*}{\textcolor{purple}{\large{\textbf{Inters.}}}}  &\textbf{random} &\textcolor{gray!65}{0.3872} &\textcolor{gray!65}{0.2854} &\textcolor{gray!65}{0.1018} &\textcolor{gray!65}{0.0509} & &\textcolor{gray!65}{0.4210} &\textcolor{gray!65}{0.3498} &\textcolor{gray!65}{0.3375} &\textcolor{gray!65}{0.0835} &\textcolor{gray!65}{0.0368} & &0.4500 &0.3719 &0.3750 &0.3381 &0.3013 &0.1875 &\colorbox{purple!30}{0.2625} &0.0807 \\
&\textbf{top-rated}  &\textcolor{gray!65}{0.8329} &\textcolor{gray!65}{0.7897} &\textcolor{gray!65}{0.0432} &\textcolor{gray!65}{0.0216} & &\textcolor{gray!65}{0.8340} &\textcolor{gray!65}{0.7970} &\textcolor{gray!65}{0.8624} &\textcolor{gray!65}{0.0654} &\textcolor{gray!65}{0.0268} & &0.8360 &0.8108 &0.8704 &0.8286 &0.7668 &0.8304 &\colorbox{purple!30}{0.1036} &0.0311 \\
&\textbf{recent}  &\textcolor{gray!65}{0.8144} &\textcolor{gray!65}{0.8164} &\textcolor{gray!65}{0.0020} &\textcolor{gray!65}{0.0010} & &\textcolor{gray!65}{0.7932} &\textcolor{gray!65}{0.8182} &\textcolor{gray!65}{0.8228} &\textcolor{gray!65}{0.0296} &\textcolor{gray!65}{0.0130} & &0.8200 &0.8093 &0.8199 &0.7168 &0.8376 &0.8345 &0.1208 &0.0412 \\

\bottomrule
\end{tabular}}

\caption*{\small{Please note that the rows represent prompting strategies, and column tabs represent group results. To test the effect of sex-based instruction prompting, we should mainly look at the results in the sex category. This is the reason the other boxes are gray. However, we will also look at the cross effect in RQ3.}}
\end{table}

\begin{table}[!h]
\caption{True preference alignment}
\label{tab:pref_align}
\centering
\renewcommand{\arraystretch}{1.3} 
\resizebox{1.05\textwidth}{!}{%
\begin{tabular}{lcccccccccccccccccccc}
\toprule

 & & \multicolumn{4}{c}{\textbf{Sex}} & & \multicolumn{5}{c}{\textbf{Age}} & & \multicolumn{7}{c}{\textbf{Intersectional (sex \& Age)}} \\
 \cline{3-6} \cline{8-12} \cline{14-21}
&  & $S_{mn}$ & $S_{fn}$ & $SNSR$ & $SNSV$ & & $S_{tn}$ & $S_{yn}$ & $S_{an}$ & $SNSR$ & $SNSV$& & $S_{mt}$ & $S_{my}$ & $S_{ma}$ & $S_{ft}$ & $S_{fy}$ & $S_{fa}$ & $SNSR$ & $SNSV$  \\
\toprule
\multicolumn{21}{c}{\large Jaccard} \\
\toprule    
 \multirow{3}{*}{\textcolor{red}{\large{\textbf{Sex}}}}  &\textbf{random} &\colorbox{purple!30}{0.0313} &\colorbox{purple!30}{0.0103} &\colorbox{purple!30}{0.0210} &0.0105 & &\textcolor{gray!65}{0.0193} &\textcolor{gray!65}{0.0271} &\textcolor{gray!65}{0.0266} &\textcolor{gray!65}{0.0078} &\textcolor{gray!65}{0.0035} & &\textcolor{gray!65}{0.0261} &\textcolor{gray!65}{0.0321} &\textcolor{gray!65}{0.0332} &\textcolor{gray!65}{0.0000} &\textcolor{gray!65}{0.0162} &\textcolor{gray!65}{0.0000} &\textcolor{gray!65}{0.0332} &\textcolor{gray!65}{0.0138} \\
 &\textbf{top-rated} &\colorbox{purple!30}{0.0681} &\colorbox{purple!30}{0.0529} &0.0152 &0.0076 & &\textcolor{gray!65}{0.0849} &\textcolor{gray!65}{0.0535} &\textcolor{gray!65}{0.0716} &\textcolor{gray!65}{0.0314} &\textcolor{gray!65}{0.0129} & &\textcolor{gray!65}{0.0846} &\textcolor{gray!65}{0.0521} &\textcolor{gray!65}{0.0863} &\textcolor{gray!65}{0.0857} &\textcolor{gray!65}{0.0565} &\textcolor{gray!65}{0.0125} &\textcolor{gray!65}{0.0738} &\textcolor{gray!65}{0.0266} \\

 &\textbf{recent} &0.0461 &0.0375 &\colorbox{purple!30}{0.0086} &0.0043 & &\textcolor{gray!65}{0.0704} &\textcolor{gray!65}{0.0398} &\textcolor{gray!65}{0.0340} &\textcolor{gray!65}{0.0365} &\textcolor{gray!65}{0.0160} & &\textcolor{gray!65}{0.0824} &\textcolor{gray!65}{0.0392} &\textcolor{gray!65}{0.0359} &\textcolor{gray!65}{0.0363} &\textcolor{gray!65}{0.0413} &\textcolor{gray!65}{0.0264} &\textcolor{gray!65}{0.0560} &\textcolor{gray!65}{0.0180} \\
\midrule
 \multirow{3}{*}{\textcolor{blue}{\large{\textbf{Age}}}}  &\textbf{random} 
&\textcolor{gray!65}{0.0189} &\textcolor{gray!65}{0.0225} &\textcolor{gray!65}{0.0036} &\textcolor{gray!65}{0.0018} & &{0.0226} &{0.0199} &{0.0181} &{0.0045} &{0.0019} & &\textcolor{gray!65}{0.0206} &\textcolor{gray!65}{0.0197} &\textcolor{gray!65}{0.0164} &\textcolor{gray!65}{0.0286} &\textcolor{gray!65}{0.0202} &\textcolor{gray!65}{0.0250} &\textcolor{gray!65}{0.0122} &\textcolor{gray!65}{0.0040} \\

&\textbf{top-rated}  &\textcolor{gray!65}{0.0640} &\textcolor{gray!65}{0.0549} &\textcolor{gray!65}{0.0091} &\textcolor{gray!65}{0.0046} & &0.0919 &0.0481 &0.0688 &0.0437 &0.0179 & &\textcolor{gray!65}{0.0890} &\textcolor{gray!65}{0.0446} &\textcolor{gray!65}{0.0829} &\textcolor{gray!65}{0.1000} &\textcolor{gray!65}{0.0558} &\textcolor{gray!65}{0.0125} &\textcolor{gray!65}{0.0875} &\textcolor{gray!65}{0.0299} \\

&\textbf{recent} &\textcolor{gray!65}{0.0487} &\textcolor{gray!65}{0.0394} &\textcolor{gray!65}{0.0093} &\textcolor{gray!65}{0.0047} & &0.0658 &0.0445 &0.0362 &0.0296 &0.0125 & &\textcolor{gray!65}{0.0771} &\textcolor{gray!65}{0.0442} &\textcolor{gray!65}{0.0390} &\textcolor{gray!65}{0.0337} &\textcolor{gray!65}{0.0453} &\textcolor{gray!65}{0.0250} &\textcolor{gray!65}{0.0521} &\textcolor{gray!65}{0.0163} \\

\midrule
\multirow{3}{*}{\textcolor{violet}{\large{\textbf{Inters.}}}}  &\textbf{random}  &\textcolor{gray!65}{0.0274} &\textcolor{gray!65}{0.0155} &\textcolor{gray!65}{0.0119} &\textcolor{gray!65}{0.0060} & &\textcolor{gray!65}{0.0335} &\textcolor{gray!65}{0.0202} &\textcolor{gray!65}{0.0260} &\textcolor{gray!65}{0.0133} &\textcolor{gray!65}{0.0055} & &0.0342 &0.0222 &0.0325 &0.0317 &0.0159 &0.0000 &0.0342 &0.0120 \\
&\textbf{top-rated} &\textcolor{gray!65}{0.0647} &\textcolor{gray!65}{0.0566} &\textcolor{gray!65}{0.0082} &\textcolor{gray!65}{0.0041} & &\textcolor{gray!65}{0.0957} &\textcolor{gray!65}{0.0519} &\textcolor{gray!65}{0.0620} &\textcolor{gray!65}{0.0438} &\textcolor{gray!65}{0.0187} & &0.0919 &0.0497 &0.0744 &0.1063 &0.0567 &0.0125 &0.0938 &0.0305 \\
&\textbf{recent} &\textcolor{gray!65}{0.0440} &\textcolor{gray!65}{0.0328} &\textcolor{gray!65}{0.0111} &\textcolor{gray!65}{0.0056} & &\textcolor{gray!65}{0.0652} &\textcolor{gray!65}{0.0365} &\textcolor{gray!65}{0.0336} &\textcolor{gray!65}{0.0316} &\textcolor{gray!65}{0.0143} & &0.0825 &0.0350 &0.0358 &0.0159 &0.0398 &0.0250 &0.0666 &0.0210 \\

\toprule
\multicolumn{21}{c}{\large PRAG} \\
\toprule
 \multirow{3}{*}{\textcolor{red}{\large{\textbf{Sex}}}}  &\textbf{random} &\colorbox{purple!30}{0.0367} &\colorbox{purple!30}{0.0000} &\colorbox{purple!30}{0.0367} &0.0183 & &\textcolor{gray!65}{0.0000} &\textcolor{gray!65}{0.0241} &\textcolor{gray!65}{0.0500} &\textcolor{gray!65}{0.0500} &\textcolor{gray!65}{0.0204} & &\textcolor{gray!65}{0.0000} &\textcolor{gray!65}{0.0351} &\textcolor{gray!65}{0.0625} &\textcolor{gray!65}{0.0000} &\textcolor{gray!65}{0.0000} &\textcolor{gray!65}{0.0000} &\textcolor{gray!65}{0.0625} &\textcolor{gray!65}{0.02437} \\
  &\textbf{top-rated} &\colorbox{purple!30}{0.0459} &\colorbox{purple!30}{0.0488} &\colorbox{purple!30}{0.0029} &0.0015 & &\textcolor{gray!65}{0.1173}  &\textcolor{gray!65}{0.0321} &\textcolor{gray!65}{0.0292} &\textcolor{gray!65}{0.0881} &\textcolor{gray!65}{0.0409} & &\textcolor{gray!65}{0.1083} &\textcolor{gray!65}{0.0292} &\textcolor{gray!65}{0.0365} &\textcolor{gray!65}{0.1429} &\textcolor{gray!65}{0.0385} &\textcolor{gray!65}{0.0000} &\textcolor{gray!65}{0.1429} &\textcolor{gray!65}{0.0496} \\
   &\textbf{recent} &0.0183 &0.0000 &\colorbox{purple!30}{0.0183} &0.0092 &&\textcolor{gray!65}{0.0741} &\textcolor{gray!65}{0.0000} &\textcolor{gray!65}{0.0000} &\textcolor{gray!65}{0.0741} &\textcolor{gray!65}{0.0349} & &\textcolor{gray!65}{0.1000} &\textcolor{gray!65}{0.0000} &\textcolor{gray!65}{0.0000} &\textcolor{gray!65}{0.0000} &\textcolor{gray!65}{0.0000} &\textcolor{gray!65}{0.0000} &\textcolor{gray!65}{0.1000} &\textcolor{gray!65}{0.0373} \\
   \midrule

 \multirow{3}{*}{\textcolor{blue}{\large{\textbf{Age}}}} 
&\textbf{random} &\textcolor{gray!65}{0.0183} &\textcolor{gray!65}{0.0244} &\textcolor{gray!65}{0.0060} &\textcolor{gray!65}{0.0030} & &0.0741 &0.0120 &0.0000 &0.0741 &0.0325 & &\textcolor{gray!65}{0.0500} &\textcolor{gray!65}{0.0175} &\textcolor{gray!65}{0.0000} &\textcolor{gray!65}{0.1429} &\textcolor{gray!65}{0.0000} &\textcolor{gray!65}{0.0000} &\textcolor{gray!65}{0.1429} &\textcolor{gray!65}{0.0514} \\

 &\textbf{top-rated} &\textcolor{gray!65}{0.0642} &\textcolor{gray!65}{0.0325} &\textcolor{gray!65}{0.0317} &\textcolor{gray!65}{0.0158} & &0.1481 &0.0361 &0.0333 &0.1148 &0.0535 & &\textcolor{gray!65}{0.1833} &\textcolor{gray!65}{0.0351} &\textcolor{gray!65}{0.0417} &\textcolor{gray!65}{0.0476} &\textcolor{gray!65}{0.0385} &\textcolor{gray!65}{0.0000} &\textcolor{gray!65}{0.1833} &\textcolor{gray!65}{0.0582} \\

 &\textbf{recent} &\textcolor{gray!65}{0.0428} &\textcolor{gray!65}{0.0244} &\textcolor{gray!65}{0.0184} &\textcolor{gray!65}{0.0092} & &0.0741 &0.0321 &0.0250 &\colorbox{purple!30}{0.0491} &0.0217 & &\textcolor{gray!65}{0.1000} &\textcolor{gray!65}{0.0292} &\textcolor{gray!65}{0.0312} &\textcolor{gray!65}{0.0000} &\textcolor{gray!65}{0.0385} &\textcolor{gray!65}{0.0000} &\textcolor{gray!65}{0.1000} &\textcolor{gray!65}{0.0335} \\

\midrule
\multirow{3}{*}{\textcolor{violet}{\large{\textbf{Inters.}}}}  &\textbf{random} &\textcolor{gray!65}{0.0092} &\textcolor{gray!65}{0.0000} &\textcolor{gray!65}{0.0092} &\textcolor{gray!65}{0.0046} & &\textcolor{gray!65}{0.0370} &\textcolor{gray!65}{0.0000} &\textcolor{gray!65}{0.0000} &\textcolor{gray!65}{0.0370} &\textcolor{gray!65}{0.0175} & &0.0500 &0.0000 &0.0000 &0.0000 &0.0000 &0.0000 &0.0500 &0.0186 \\
 &\textbf{top-rated} &\textcolor{gray!65}{0.0443} &\textcolor{gray!65}{0.0366} &\textcolor{gray!65}{0.0078} &\textcolor{gray!65}{0.0039} & &\textcolor{gray!65}{0.0988} &\textcolor{gray!65}{0.0361} &\textcolor{gray!65}{0.0167} &\textcolor{gray!65}{0.0821} &\textcolor{gray!65}{0.0350} & &0.1083 &0.0351 &0.0208 &0.0714 &0.0385 &0.0000 &0.1083 &0.0353 \\
 &\textbf{recent} &\textcolor{gray!65}{0.0183} &\textcolor{gray!65}{0.0000} &\textcolor{gray!65}{0.0183} &\textcolor{gray!65}{0.0092} & &\textcolor{gray!65}{0.0370} &\textcolor{gray!65}{0.0000} &\textcolor{gray!65}{0.0250} &\textcolor{gray!65}{0.0370} &\textcolor{gray!65}{0.0154} & &0.0500 &0.0000 &0.0312 &0.0000 &0.0000 &0.0000 &\colorbox{purple!30}{0.0500} &0.0199 \\
\bottomrule
\end{tabular}}
\caption*{\small{Please note that the rows represent prompting strategies, and column tabs represent group results. To test the effect of sex-based instruction prompting, we should mainly look at the results in the sex category. This is the reason the other boxes are gray. However, we will also look at the cross effect in RQ3.}}
\end{table}

\bibliographystyle{ACM-Reference-Format}
\bibliography{refs}


\begin{thebibliography}{67}


\ifx \showCODEN    \undefined \def \showCODEN     #1{\unskip}     \fi
\ifx \showDOI      \undefined \def \showDOI       #1{#1}\fi
\ifx \showISBNx    \undefined \def \showISBNx     #1{\unskip}     \fi
\ifx \showISBNxiii \undefined \def \showISBNxiii  #1{\unskip}     \fi
\ifx \showISSN     \undefined \def \showISSN      #1{\unskip}     \fi
\ifx \showLCCN     \undefined \def \showLCCN      #1{\unskip}     \fi
\ifx \shownote     \undefined \def \shownote      #1{#1}          \fi
\ifx \showarticletitle \undefined \def \showarticletitle #1{#1}   \fi
\ifx \showURL      \undefined \def \showURL       {\relax}        \fi
\providecommand\bibfield[2]{#2}
\providecommand\bibinfo[2]{#2}
\providecommand\natexlab[1]{#1}
\providecommand\showeprint[2][]{arXiv:#2}

\bibitem[Amig{\'o} et~al\mbox{.}(2023)]%
        {deldjoo2021explaining}
\bibfield{author}{\bibinfo{person}{Enrique Amig{\'o}}, \bibinfo{person}{Yashar Deldjoo}, \bibinfo{person}{Stefano Mizzaro}, {and} \bibinfo{person}{Alejandro Bellog{\'\i}n}.} \bibinfo{year}{2023}\natexlab{}.
\newblock \showarticletitle{A unifying and general account of fairness measurement in recommender systems}.
\newblock \bibinfo{journal}{\emph{Information Processing \& Management}} \bibinfo{volume}{60}, \bibinfo{number}{1} (\bibinfo{year}{2023}), \bibinfo{pages}{103115}.
\newblock


\bibitem[Asai et~al\mbox{.}(2022)]%
        {asai2022attempt}
\bibfield{author}{\bibinfo{person}{Akari Asai}, \bibinfo{person}{Mohammadreza Salehi}, \bibinfo{person}{Matthew~E Peters}, {and} \bibinfo{person}{Hannaneh Hajishirzi}.} \bibinfo{year}{2022}\natexlab{}.
\newblock \showarticletitle{Attempt: Parameter-efficient multi-task tuning via attentional mixtures of soft prompts}. In \bibinfo{booktitle}{\emph{Proceedings of the 2022 Conference on Empirical Methods in Natural Language Processing}}. \bibinfo{pages}{6655--6672}.
\newblock


\bibitem[Biancofiore et~al\mbox{.}(2024)]%
        {biancofiore2024interactive}
\bibfield{author}{\bibinfo{person}{Giovanni~Maria Biancofiore}, \bibinfo{person}{Yashar Deldjoo}, \bibinfo{person}{Tommaso Di~Noia}, \bibinfo{person}{Eugenio Di~Sciascio}, {and} \bibinfo{person}{Fedelucio Narducci}.} \bibinfo{year}{2024}\natexlab{}.
\newblock \showarticletitle{Interactive Question Answering Systems: Literature Review}.
\newblock \bibinfo{journal}{\emph{ACM Computing Surveys (CSUR)}} (\bibinfo{year}{2024}).
\newblock


\bibitem[Bogers and Koolen(2017)]%
        {bogers2017defining}
\bibfield{author}{\bibinfo{person}{Toine Bogers} {and} \bibinfo{person}{Marijn Koolen}.} \bibinfo{year}{2017}\natexlab{}.
\newblock \showarticletitle{Defining and supporting narrative-driven recommendation}. In \bibinfo{booktitle}{\emph{Proceedings of the eleventh ACM conference on recommender systems}}. \bibinfo{pages}{238--242}.
\newblock


\bibitem[Boratto et~al\mbox{.}(2021)]%
        {boratto2021interplay}
\bibfield{author}{\bibinfo{person}{Ludovico Boratto}, \bibinfo{person}{Gianni Fenu}, {and} \bibinfo{person}{Mirko Marras}.} \bibinfo{year}{2021}\natexlab{}.
\newblock \showarticletitle{Interplay between upsampling and regularization for provider fairness in recommender systems}.
\newblock \bibinfo{journal}{\emph{User Modeling and User-Adapted Interaction}} \bibinfo{volume}{31}, \bibinfo{number}{3} (\bibinfo{year}{2021}), \bibinfo{pages}{421--455}.
\newblock


\bibitem[Brown et~al\mbox{.}(2020)]%
        {brown2020language}
\bibfield{author}{\bibinfo{person}{Tom Brown}, \bibinfo{person}{Benjamin Mann}, \bibinfo{person}{Nick Ryder}, \bibinfo{person}{Melanie Subbiah}, \bibinfo{person}{Jared~D Kaplan}, \bibinfo{person}{Prafulla Dhariwal}, \bibinfo{person}{Arvind Neelakantan}, \bibinfo{person}{Pranav Shyam}, \bibinfo{person}{Girish Sastry}, \bibinfo{person}{Amanda Askell}, {et~al\mbox{.}}} \bibinfo{year}{2020}\natexlab{}.
\newblock \showarticletitle{Language models are few-shot learners}.
\newblock \bibinfo{journal}{\emph{Advances in neural information processing systems}}  \bibinfo{volume}{33} (\bibinfo{year}{2020}), \bibinfo{pages}{1877--1901}.
\newblock


\bibitem[Burke et~al\mbox{.}(2018)]%
        {burke2018balanced}
\bibfield{author}{\bibinfo{person}{Robin Burke}, \bibinfo{person}{Nasim Sonboli}, {and} \bibinfo{person}{Aldo Ordonez-Gauger}.} \bibinfo{year}{2018}\natexlab{}.
\newblock \showarticletitle{Balanced neighborhoods for multi-sided fairness in recommendation}. In \bibinfo{booktitle}{\emph{Conference on fairness, accountability and transparency}}. PMLR, \bibinfo{pages}{202--214}.
\newblock


\bibitem[Chakraborty et~al\mbox{.}(2017)]%
        {chakraborty2017fair}
\bibfield{author}{\bibinfo{person}{Abhijnan Chakraborty}, \bibinfo{person}{Aniko Hannak}, \bibinfo{person}{Asia~J Biega}, {and} \bibinfo{person}{Krishna~P Gummadi}.} \bibinfo{year}{2017}\natexlab{}.
\newblock \showarticletitle{Fair sharing for sharing economy platforms}.
\newblock  (\bibinfo{year}{2017}).
\newblock


\bibitem[Chang et~al\mbox{.}(2023)]%
        {chang2023survey}
\bibfield{author}{\bibinfo{person}{Yupeng Chang}, \bibinfo{person}{Xu Wang}, \bibinfo{person}{Jindong Wang}, \bibinfo{person}{Yuan Wu}, \bibinfo{person}{Linyi Yang}, \bibinfo{person}{Kaijie Zhu}, \bibinfo{person}{Hao Chen}, \bibinfo{person}{Xiaoyuan Yi}, \bibinfo{person}{Cunxiang Wang}, \bibinfo{person}{Yidong Wang}, {et~al\mbox{.}}} \bibinfo{year}{2023}\natexlab{}.
\newblock \showarticletitle{A survey on evaluation of large language models}.
\newblock \bibinfo{journal}{\emph{ACM Transactions on Intelligent Systems and Technology}} (\bibinfo{year}{2023}).
\newblock


\bibitem[da~Silva et~al\mbox{.}(2021)]%
        {da2021exploiting}
\bibfield{author}{\bibinfo{person}{Diego~Corr{\^e}a da Silva}, \bibinfo{person}{Marcelo~Garcia Manzato}, {and} \bibinfo{person}{Frederico~Ara{\'u}jo Dur{\~a}o}.} \bibinfo{year}{2021}\natexlab{}.
\newblock \showarticletitle{Exploiting personalized calibration and metrics for fairness recommendation}.
\newblock \bibinfo{journal}{\emph{Expert Systems with Applications}}  \bibinfo{volume}{181} (\bibinfo{year}{2021}), \bibinfo{pages}{115112}.
\newblock


\bibitem[De~Gemmis et~al\mbox{.}(2015)]%
        {de2015semantics}
\bibfield{author}{\bibinfo{person}{Marco De~Gemmis}, \bibinfo{person}{Pasquale Lops}, \bibinfo{person}{Cataldo Musto}, \bibinfo{person}{Fedelucio Narducci}, {and} \bibinfo{person}{Giovanni Semeraro}.} \bibinfo{year}{2015}\natexlab{}.
\newblock \showarticletitle{Semantics-aware content-based recommender systems}.
\newblock \bibinfo{journal}{\emph{Recommender systems handbook}} (\bibinfo{year}{2015}), \bibinfo{pages}{119--159}.
\newblock


\bibitem[Deldjoo(2023)]%
        {deldjoo2023fairnessgpt}
\bibfield{author}{\bibinfo{person}{Yashar Deldjoo}.} \bibinfo{year}{2023}\natexlab{}.
\newblock \showarticletitle{Fairness of ChatGPT and the Role of Explainable-Guided Prompts}. In \bibinfo{booktitle}{\emph{COLLM@ECML-PKDD'23}}.
\newblock


\bibitem[Deldjoo(2024)]%
        {deldjoo2024understanding}
\bibfield{author}{\bibinfo{person}{Yashar Deldjoo}.} \bibinfo{year}{2024}\natexlab{}.
\newblock \showarticletitle{Understanding Biases in ChatGPT-based Recommender Systems: Provider Fairness, Temporal stability, and Recency}.
\newblock \bibinfo{journal}{\emph{ACM Transactions on Recommender Systems}} (\bibinfo{year}{2024}).
\newblock


\bibitem[Deldjoo et~al\mbox{.}(2021)]%
        {deldjoo2021flexible}
\bibfield{author}{\bibinfo{person}{Yashar Deldjoo}, \bibinfo{person}{Vito~Walter Anelli}, \bibinfo{person}{Hamed Zamani}, \bibinfo{person}{Alejandro Bellogin}, {and} \bibinfo{person}{Tommaso Di~Noia}.} \bibinfo{year}{2021}\natexlab{}.
\newblock \showarticletitle{A flexible framework for evaluating user and item fairness in recommender systems}.
\newblock \bibinfo{journal}{\emph{User Modeling and User-Adapted Interaction}} (\bibinfo{year}{2021}), \bibinfo{pages}{1--55}.
\newblock


\bibitem[Deldjoo et~al\mbox{.}(2020)]%
        {deldjoo2020dataset}
\bibfield{author}{\bibinfo{person}{Yashar Deldjoo}, \bibinfo{person}{Tommaso Di~Noia}, \bibinfo{person}{Eugenio Di~Sciascio}, {and} \bibinfo{person}{Felice~Antonio Merra}.} \bibinfo{year}{2020}\natexlab{}.
\newblock \showarticletitle{How Dataset Characteristics Affect the Robustness of Collaborative Recommendation Models}. In \bibinfo{booktitle}{\emph{Proceedings of the 43rd International ACM SIGIR Conference on Research and Development in Information Retrieval}}.
\newblock


\bibitem[Deldjoo et~al\mbox{.}(2022)]%
        {deldjoo2022survey}
\bibfield{author}{\bibinfo{person}{Yashar Deldjoo}, \bibinfo{person}{Tommaso Di~Noia}, {and} \bibinfo{person}{Felice~Antonio Merra}.} \bibinfo{year}{2022}\natexlab{}.
\newblock \showarticletitle{A survey on adversarial recommender systems: from attack/defense strategies to generative adversarial networks}.
\newblock \bibinfo{journal}{\emph{Comput. Surveys}} \bibinfo{number}{2} (\bibinfo{year}{2022}), \bibinfo{pages}{1--38}.
\newblock


\bibitem[Deldjoo et~al\mbox{.}(2024a)]%
        {deldjoo2024review}
\bibfield{author}{\bibinfo{person}{Yashar Deldjoo}, \bibinfo{person}{Zhankui He}, \bibinfo{person}{Julian McAuley}, \bibinfo{person}{Anton Korikov}, \bibinfo{person}{Scott Sanner}, \bibinfo{person}{Arnau Ramisa}, \bibinfo{person}{Ren{\'e} Vidal}, \bibinfo{person}{Maheswaran Sathiamoorthy}, \bibinfo{person}{Atoosa Kasirzadeh}, {and} \bibinfo{person}{Silvia Milano}.} \bibinfo{year}{2024}\natexlab{a}.
\newblock \showarticletitle{A Review of Modern Recommender Systems using Generative Models (Gen-RecSys)}. In \bibinfo{booktitle}{\emph{Proceedings of the 30th ACM SIGKDD Conference on Knowledge Discovery and Data Mining}}. \bibinfo{pages}{6448--6458}.
\newblock


\bibitem[Deldjoo et~al\mbox{.}(2024b)]%
        {deldjoo2024recommendation}
\bibfield{author}{\bibinfo{person}{Yashar Deldjoo}, \bibinfo{person}{Zhankui He}, \bibinfo{person}{Julian McAuley}, \bibinfo{person}{Anton Korikov}, \bibinfo{person}{Scott Sanner}, \bibinfo{person}{Arnau Ramisa}, \bibinfo{person}{Rene Vidal}, \bibinfo{person}{Maheswaran Sathiamoorthy}, \bibinfo{person}{Atoosa Kasrizadeh}, \bibinfo{person}{Silvia Milano}, {et~al\mbox{.}}} \bibinfo{year}{2024}\natexlab{b}.
\newblock \showarticletitle{Recommendation with generative models}.
\newblock \bibinfo{journal}{\emph{arXiv preprint arXiv:2409.15173}} (\bibinfo{year}{2024}).
\newblock


\bibitem[Deldjoo et~al\mbox{.}(2023)]%
        {deldjoo2023fairness}
\bibfield{author}{\bibinfo{person}{Yashar Deldjoo}, \bibinfo{person}{Dietmar Jannach}, \bibinfo{person}{Alejandro Bellogin}, \bibinfo{person}{Alessandro Difonzo}, {and} \bibinfo{person}{Dario Zanzonelli}.} \bibinfo{year}{2023}\natexlab{}.
\newblock \showarticletitle{Fairness in recommender systems: research landscape and future directions}.
\newblock \bibinfo{journal}{\emph{User Modeling and User-Adapted Interaction}} (\bibinfo{year}{2023}), \bibinfo{pages}{1--50}.
\newblock


\bibitem[Deldjoo and Nazary(2024)]%
        {deldjoo2024normative}
\bibfield{author}{\bibinfo{person}{Yashar Deldjoo} {and} \bibinfo{person}{Fatemeh Nazary}.} \bibinfo{year}{2024}\natexlab{}.
\newblock \showarticletitle{A Normative Framework for Benchmarking Consumer Fairness in Large Language Model Recommender System}. In \bibinfo{booktitle}{\emph{ROEGen@RecSys'24}}.
\newblock


\bibitem[Deldjoo et~al\mbox{.}(2018)]%
        {deldjoo2018content}
\bibfield{author}{\bibinfo{person}{Yashar Deldjoo}, \bibinfo{person}{Markus Schedl}, \bibinfo{person}{Paolo Cremonesi}, \bibinfo{person}{Gabirella Pasi}, {et~al\mbox{.}}} \bibinfo{year}{2018}\natexlab{}.
\newblock \showarticletitle{Content-based multimedia recommendation systems: definition and application domains}. In \bibinfo{booktitle}{\emph{Italian Information Retrieval Workshop}}. \bibinfo{pages}{1--4}.
\newblock


\bibitem[Di~Palma et~al\mbox{.}(2023)]%
        {di2023evaluating}
\bibfield{author}{\bibinfo{person}{Dario Di~Palma}, \bibinfo{person}{Giovanni~Maria Biancofiore}, \bibinfo{person}{Vito~Walter Anelli}, \bibinfo{person}{Fedelucio Narducci}, \bibinfo{person}{Tommaso Di~Noia}, {and} \bibinfo{person}{Eugenio Di~Sciascio}.} \bibinfo{year}{2023}\natexlab{}.
\newblock \showarticletitle{Evaluating chatgpt as a recommender system: A rigorous approach}.
\newblock \bibinfo{journal}{\emph{arXiv preprint arXiv:2309.03613}} (\bibinfo{year}{2023}).
\newblock


\bibitem[Do et~al\mbox{.}(2021)]%
        {do2021two}
\bibfield{author}{\bibinfo{person}{Virginie Do}, \bibinfo{person}{Sam Corbett-Davies}, \bibinfo{person}{Jamal Atif}, {and} \bibinfo{person}{Nicolas Usunier}.} \bibinfo{year}{2021}\natexlab{}.
\newblock \showarticletitle{Two-sided fairness in rankings via Lorenz dominance}.
\newblock \bibinfo{journal}{\emph{Advances in Neural Information Processing Systems}}  \bibinfo{volume}{34} (\bibinfo{year}{2021}).
\newblock


\bibitem[Dong et~al\mbox{.}(2021)]%
        {dong2021user}
\bibfield{author}{\bibinfo{person}{Qiang Dong}, \bibinfo{person}{Shuang-Shuang Xie}, {and} \bibinfo{person}{Wen-Jun Li}.} \bibinfo{year}{2021}\natexlab{}.
\newblock \showarticletitle{User-item matching for recommendation fairness}.
\newblock \bibinfo{journal}{\emph{IEEE Access}}  \bibinfo{volume}{9} (\bibinfo{year}{2021}), \bibinfo{pages}{130389--130398}.
\newblock


\bibitem[Dowling and Lucey(2023)]%
        {dowling2023chatgpt}
\bibfield{author}{\bibinfo{person}{Michael Dowling} {and} \bibinfo{person}{Brian Lucey}.} \bibinfo{year}{2023}\natexlab{}.
\newblock \showarticletitle{ChatGPT for (finance) research: The Bananarama conjecture}.
\newblock \bibinfo{journal}{\emph{Finance Research Letters}}  \bibinfo{volume}{53} (\bibinfo{year}{2023}), \bibinfo{pages}{103662}.
\newblock


\bibitem[Ekstrand et~al\mbox{.}(2019)]%
        {ekstrand2019fairness}
\bibfield{author}{\bibinfo{person}{Michael~D Ekstrand}, \bibinfo{person}{Robin Burke}, {and} \bibinfo{person}{Fernando Diaz}.} \bibinfo{year}{2019}\natexlab{}.
\newblock \showarticletitle{Fairness and discrimination in recommendation and retrieval}. In \bibinfo{booktitle}{\emph{Proceedings of the 13th ACM Conference on Recommender Systems}}. \bibinfo{pages}{576--577}.
\newblock


\bibitem[Farnadi et~al\mbox{.}(2018)]%
        {farnadi2018fairness}
\bibfield{author}{\bibinfo{person}{Golnoosh Farnadi}, \bibinfo{person}{Pigi Kouki}, \bibinfo{person}{Spencer~K Thompson}, \bibinfo{person}{Sriram Srinivasan}, {and} \bibinfo{person}{Lise Getoor}.} \bibinfo{year}{2018}\natexlab{}.
\newblock \showarticletitle{A fairness-aware hybrid recommender system}.
\newblock \bibinfo{journal}{\emph{arXiv preprint arXiv:1809.09030}} (\bibinfo{year}{2018}).
\newblock


\bibitem[Ge et~al\mbox{.}(2021)]%
        {ge2021towards}
\bibfield{author}{\bibinfo{person}{Yingqiang Ge}, \bibinfo{person}{Shuchang Liu}, \bibinfo{person}{Ruoyuan Gao}, \bibinfo{person}{Yikun Xian}, \bibinfo{person}{Yunqi Li}, \bibinfo{person}{Xiangyu Zhao}, \bibinfo{person}{Changhua Pei}, \bibinfo{person}{Fei Sun}, \bibinfo{person}{Junfeng Ge}, \bibinfo{person}{Wenwu Ou}, {et~al\mbox{.}}} \bibinfo{year}{2021}\natexlab{}.
\newblock \showarticletitle{Towards long-term fairness in recommendation}. In \bibinfo{booktitle}{\emph{Proceedings of the 14th ACM international conference on web search and data mining}}. \bibinfo{pages}{445--453}.
\newblock


\bibitem[Geng et~al\mbox{.}(2022)]%
        {geng2022recommendation}
\bibfield{author}{\bibinfo{person}{Shijie Geng}, \bibinfo{person}{Shuchang Liu}, \bibinfo{person}{Zuohui Fu}, \bibinfo{person}{Yingqiang Ge}, {and} \bibinfo{person}{Yongfeng Zhang}.} \bibinfo{year}{2022}\natexlab{}.
\newblock \showarticletitle{Recommendation as language processing (rlp): A unified pretrain, personalized prompt \& predict paradigm (p5)}. In \bibinfo{booktitle}{\emph{Proceedings of the 16th ACM Conference on Recommender Systems}}. \bibinfo{pages}{299--315}.
\newblock


\bibitem[G{\'o}mez et~al\mbox{.}(2021)]%
        {gomez2021winner}
\bibfield{author}{\bibinfo{person}{Elizabeth G{\'o}mez}, \bibinfo{person}{Carlos Shui~Zhang}, \bibinfo{person}{Ludovico Boratto}, \bibinfo{person}{Maria Salam{\'o}}, {and} \bibinfo{person}{Mirko Marras}.} \bibinfo{year}{2021}\natexlab{}.
\newblock \showarticletitle{The winner takes it all: geographic imbalance and provider (un) fairness in educational recommender systems}. In \bibinfo{booktitle}{\emph{Proceedings of the 44th International ACM SIGIR Conference on Research and Development in Information Retrieval}}. \bibinfo{pages}{1808--1812}.
\newblock


\bibitem[Hao et~al\mbox{.}(2021)]%
        {hao2021pareto}
\bibfield{author}{\bibinfo{person}{Qianxiu Hao}, \bibinfo{person}{Qianqian Xu}, \bibinfo{person}{Zhiyong Yang}, {and} \bibinfo{person}{Qingming Huang}.} \bibinfo{year}{2021}\natexlab{}.
\newblock \showarticletitle{Pareto optimality for fairness-constrained collaborative filtering}. In \bibinfo{booktitle}{\emph{Proceedings of the 29th ACM International Conference on Multimedia}}. \bibinfo{pages}{5619--5627}.
\newblock


\bibitem[He et~al\mbox{.}(2023)]%
        {he2023large}
\bibfield{author}{\bibinfo{person}{Zhankui He}, \bibinfo{person}{Zhouhang Xie}, \bibinfo{person}{Rahul Jha}, \bibinfo{person}{Harald Steck}, \bibinfo{person}{Dawen Liang}, \bibinfo{person}{Yesu Feng}, \bibinfo{person}{Bodhisattwa~Prasad Majumder}, \bibinfo{person}{Nathan Kallus}, {and} \bibinfo{person}{Julian McAuley}.} \bibinfo{year}{2023}\natexlab{}.
\newblock \showarticletitle{Large language models as zero-shot conversational recommenders}. In \bibinfo{booktitle}{\emph{Proceedings of the 32nd ACM international conference on information and knowledge management}}. \bibinfo{pages}{720--730}.
\newblock


\bibitem[Hou et~al\mbox{.}(2022)]%
        {hou2022towards}
\bibfield{author}{\bibinfo{person}{Yupeng Hou}, \bibinfo{person}{Shanlei Mu}, \bibinfo{person}{Wayne~Xin Zhao}, \bibinfo{person}{Yaliang Li}, \bibinfo{person}{Bolin Ding}, {and} \bibinfo{person}{Ji-Rong Wen}.} \bibinfo{year}{2022}\natexlab{}.
\newblock \showarticletitle{Towards universal sequence representation learning for recommender systems}. In \bibinfo{booktitle}{\emph{Proceedings of the 28th ACM SIGKDD Conference on Knowledge Discovery and Data Mining}}. \bibinfo{pages}{585--593}.
\newblock


\bibitem[Jin et~al\mbox{.}(2024)]%
        {jin2024health}
\bibfield{author}{\bibinfo{person}{Mingyu Jin}, \bibinfo{person}{Qinkai Yu}, \bibinfo{person}{Chong Zhang}, \bibinfo{person}{Dong Shu}, \bibinfo{person}{Suiyuan Zhu}, \bibinfo{person}{Mengnan Du}, \bibinfo{person}{Yongfeng Zhang}, {and} \bibinfo{person}{Yanda Meng}.} \bibinfo{year}{2024}\natexlab{}.
\newblock \showarticletitle{Health-LLM: Personalized Retrieval-Augmented Disease Prediction Model}.
\newblock \bibinfo{journal}{\emph{arXiv preprint arXiv:2402.00746}} (\bibinfo{year}{2024}).
\newblock


\bibitem[Kattnig et~al\mbox{.}(2024)]%
        {kattnig2024assessing}
\bibfield{author}{\bibinfo{person}{Markus Kattnig}, \bibinfo{person}{Alessa Angerschmid}, \bibinfo{person}{Thomas Reichel}, {and} \bibinfo{person}{Roman Kern}.} \bibinfo{year}{2024}\natexlab{}.
\newblock \showarticletitle{Assessing trustworthy AI: Technical and legal perspectives of fairness in AI}.
\newblock \bibinfo{journal}{\emph{Computer Law \& Security Review}}  \bibinfo{volume}{55} (\bibinfo{year}{2024}), \bibinfo{pages}{106053}.
\newblock


\bibitem[K{\i}rnap et~al\mbox{.}(2021)]%
        {kirnap2021estimation}
\bibfield{author}{\bibinfo{person}{{\"O}mer K{\i}rnap}, \bibinfo{person}{Fernando Diaz}, \bibinfo{person}{Asia Biega}, \bibinfo{person}{Michael Ekstrand}, \bibinfo{person}{Ben Carterette}, {and} \bibinfo{person}{Emine Yilmaz}.} \bibinfo{year}{2021}\natexlab{}.
\newblock \showarticletitle{Estimation of fair ranking metrics with incomplete judgments}. In \bibinfo{booktitle}{\emph{Proceedings of the Web Conference 2021}}. \bibinfo{pages}{1065--1075}.
\newblock


\bibitem[Lee and Lee(2015)]%
        {lee2015escaping}
\bibfield{author}{\bibinfo{person}{Kibeom Lee} {and} \bibinfo{person}{Kyogu Lee}.} \bibinfo{year}{2015}\natexlab{}.
\newblock \showarticletitle{Escaping your comfort zone: A graph-based recommender system for finding novel recommendations among relevant items}.
\newblock \bibinfo{journal}{\emph{Expert Systems with Applications}} \bibinfo{volume}{42}, \bibinfo{number}{10} (\bibinfo{year}{2015}), \bibinfo{pages}{4851--4858}.
\newblock


\bibitem[Li et~al\mbox{.}(2023b)]%
        {li2023large}
\bibfield{author}{\bibinfo{person}{Lei Li}, \bibinfo{person}{Yongfeng Zhang}, \bibinfo{person}{Dugang Liu}, {and} \bibinfo{person}{Li Chen}.} \bibinfo{year}{2023}\natexlab{b}.
\newblock \showarticletitle{Large language models for generative recommendation: A survey and visionary discussions}.
\newblock \bibinfo{journal}{\emph{arXiv preprint arXiv:2309.01157}} (\bibinfo{year}{2023}).
\newblock


\bibitem[Li et~al\mbox{.}(2023a)]%
        {li2023preliminary}
\bibfield{author}{\bibinfo{person}{Xinyi Li}, \bibinfo{person}{Yongfeng Zhang}, {and} \bibinfo{person}{Edward~C Malthouse}.} \bibinfo{year}{2023}\natexlab{a}.
\newblock \showarticletitle{A Preliminary Study of ChatGPT on News Recommendation: Personalization, Provider Fairness, Fake News}.
\newblock \bibinfo{journal}{\emph{arXiv preprint arXiv:2306.10702}} (\bibinfo{year}{2023}).
\newblock


\bibitem[Li et~al\mbox{.}(2021)]%
        {li2021user}
\bibfield{author}{\bibinfo{person}{Yunqi Li}, \bibinfo{person}{Hanxiong Chen}, \bibinfo{person}{Zuohui Fu}, \bibinfo{person}{Yingqiang Ge}, {and} \bibinfo{person}{Yongfeng Zhang}.} \bibinfo{year}{2021}\natexlab{}.
\newblock \showarticletitle{User-oriented fairness in recommendation}. In \bibinfo{booktitle}{\emph{Proceedings of the Web Conference 2021}}. \bibinfo{pages}{624--632}.
\newblock


\bibitem[Liao et~al\mbox{.}(2023)]%
        {liao2023proactive}
\bibfield{author}{\bibinfo{person}{Lizi Liao}, \bibinfo{person}{Grace~Hui Yang}, {and} \bibinfo{person}{Chirag Shah}.} \bibinfo{year}{2023}\natexlab{}.
\newblock \showarticletitle{Proactive conversational agents in the post-chatgpt world}. In \bibinfo{booktitle}{\emph{Proceedings of the 46th International ACM SIGIR Conference on Research and Development in Information Retrieval}}. \bibinfo{pages}{3452--3455}.
\newblock


\bibitem[Lin et~al\mbox{.}(2021)]%
        {lin2021mitigating}
\bibfield{author}{\bibinfo{person}{Chen Lin}, \bibinfo{person}{Xinyi Liu}, \bibinfo{person}{Guipeng Xv}, {and} \bibinfo{person}{Hui Li}.} \bibinfo{year}{2021}\natexlab{}.
\newblock \showarticletitle{Mitigating sentiment bias for recommender systems}. In \bibinfo{booktitle}{\emph{Proceedings of the 44th International ACM SIGIR Conference on Research and Development in Information Retrieval}}. \bibinfo{pages}{31--40}.
\newblock


\bibitem[Liu et~al\mbox{.}(2019)]%
        {liu2019advertisement}
\bibfield{author}{\bibinfo{person}{Duen-Ren Liu}, \bibinfo{person}{Yu-Shan Liao}, \bibinfo{person}{Ya-Han Chung}, {and} \bibinfo{person}{Kuan-Yu Chen}.} \bibinfo{year}{2019}\natexlab{}.
\newblock \showarticletitle{Advertisement recommendation based on personal interests and ad push fairness}.
\newblock \bibinfo{journal}{\emph{Kybernetes}} \bibinfo{volume}{48}, \bibinfo{number}{8} (\bibinfo{year}{2019}), \bibinfo{pages}{1586--1605}.
\newblock


\bibitem[Liu et~al\mbox{.}(2020)]%
        {liu2020balancing}
\bibfield{author}{\bibinfo{person}{Weiwen Liu}, \bibinfo{person}{Feng Liu}, \bibinfo{person}{Ruiming Tang}, \bibinfo{person}{Ben Liao}, \bibinfo{person}{Guangyong Chen}, {and} \bibinfo{person}{Pheng~Ann Heng}.} \bibinfo{year}{2020}\natexlab{}.
\newblock \showarticletitle{Balancing between accuracy and fairness for interactive recommendation with reinforcement learning}. In \bibinfo{booktitle}{\emph{Advances in Knowledge Discovery and Data Mining: 24th Pacific-Asia Conference, PAKDD 2020, Singapore, May 11--14, 2020, Proceedings, Part I 24}}. Springer, \bibinfo{pages}{155--167}.
\newblock


\bibitem[Naghiaei et~al\mbox{.}(2022)]%
        {naghiaei2022cpfair}
\bibfield{author}{\bibinfo{person}{Mohammadmehdi Naghiaei}, \bibinfo{person}{Hossein~A Rahmani}, {and} \bibinfo{person}{Yashar Deldjoo}.} \bibinfo{year}{2022}\natexlab{}.
\newblock \showarticletitle{Cpfair: Personalized consumer and producer fairness re-ranking for recommender systems}. In \bibinfo{booktitle}{\emph{Proceedings of the 45th International ACM SIGIR Conference on Research and Development in Information Retrieval}}. \bibinfo{pages}{770--779}.
\newblock


\bibitem[Nazary et~al\mbox{.}(2023)]%
        {nazary2023chatgpt}
\bibfield{author}{\bibinfo{person}{Fatemeh Nazary}, \bibinfo{person}{Yashar Deldjoo}, {and} \bibinfo{person}{Tommaso Di~Noia}.} \bibinfo{year}{2023}\natexlab{}.
\newblock \showarticletitle{ChatGPT-HealthPrompt. Harnessing the Power of XAI in Prompt-Based Healthcare Decision Support using ChatGPT}. In \bibinfo{booktitle}{\emph{European Conference on Artificial Intelligence}}. Springer, \bibinfo{pages}{382--397}.
\newblock


\bibitem[Nazary et~al\mbox{.}(2025)]%
        {nazary2025poison}
\bibfield{author}{\bibinfo{person}{Fatemeh Nazary}, \bibinfo{person}{Yashar Deldjoo}, {and} \bibinfo{person}{Tommaso di Noia}.} \bibinfo{year}{2025}\natexlab{}.
\newblock \showarticletitle{Poison-RAG: Adversarial Data Poisoning Attacks on Retrieval-Augmented Generation in Recommender Systems}.
\newblock \bibinfo{journal}{\emph{arXiv preprint arXiv:2501.11759}} (\bibinfo{year}{2025}).
\newblock


\bibitem[Nazary et~al\mbox{.}(2024)]%
        {nazary2024xai4llm}
\bibfield{author}{\bibinfo{person}{Fatemeh Nazary}, \bibinfo{person}{Yashar Deldjoo}, \bibinfo{person}{Tommaso Di~Noia}, {and} \bibinfo{person}{Eugenio di Sciascio}.} \bibinfo{year}{2024}\natexlab{}.
\newblock \showarticletitle{XAI4LLM. Let Machine Learning Models and LLMs Collaborate for Enhanced In-Context Learning in Healthcare}.
\newblock \bibinfo{journal}{\emph{arXiv preprint arXiv:2405.06270}} (\bibinfo{year}{2024}).
\newblock


\bibitem[Patro et~al\mbox{.}(2020)]%
        {patro2020fairrec}
\bibfield{author}{\bibinfo{person}{Gourab~K Patro}, \bibinfo{person}{Arpita Biswas}, \bibinfo{person}{Niloy Ganguly}, \bibinfo{person}{Krishna~P Gummadi}, {and} \bibinfo{person}{Abhijnan Chakraborty}.} \bibinfo{year}{2020}\natexlab{}.
\newblock \showarticletitle{Fairrec: Two-sided fairness for personalized recommendations in two-sided platforms}. In \bibinfo{booktitle}{\emph{Proceedings of The Web Conference 2020}}. \bibinfo{pages}{1194--1204}.
\newblock


\bibitem[Rahmani et~al\mbox{.}(2022)]%
        {rahmani2022unfairness}
\bibfield{author}{\bibinfo{person}{Hossein~A Rahmani}, \bibinfo{person}{Yashar Deldjoo}, \bibinfo{person}{Ali Tourani}, {and} \bibinfo{person}{Mohammadmehdi Naghiaei}.} \bibinfo{year}{2022}\natexlab{}.
\newblock \showarticletitle{The Unfairness of Active Users and Popularity Bias in Point-of-Interest Recommendation}. In \bibinfo{booktitle}{\emph{Bias@ECIR'22}}.
\newblock


\bibitem[Sanner et~al\mbox{.}(2023)]%
        {sanner2023large}
\bibfield{author}{\bibinfo{person}{Scott Sanner}, \bibinfo{person}{Krisztian Balog}, \bibinfo{person}{Filip Radlinski}, \bibinfo{person}{Ben Wedin}, {and} \bibinfo{person}{Lucas Dixon}.} \bibinfo{year}{2023}\natexlab{}.
\newblock \showarticletitle{Large language models are competitive near cold-start recommenders for language-and item-based preferences}. In \bibinfo{booktitle}{\emph{Proceedings of the 17th ACM conference on recommender systems}}. \bibinfo{pages}{890--896}.
\newblock


\bibitem[Shakespeare et~al\mbox{.}(2020)]%
        {shakespeare2020exploring}
\bibfield{author}{\bibinfo{person}{Dougal Shakespeare}, \bibinfo{person}{Lorenzo Porcaro}, \bibinfo{person}{Emilia G{\'o}mez}, {and} \bibinfo{person}{Carlos Castillo}.} \bibinfo{year}{2020}\natexlab{}.
\newblock \showarticletitle{Exploring artist gender bias in music recommendation}.
\newblock \bibinfo{journal}{\emph{arXiv preprint arXiv:2009.01715}} (\bibinfo{year}{2020}).
\newblock


\bibitem[Shen et~al\mbox{.}(2023)]%
        {shen2023towards}
\bibfield{author}{\bibinfo{person}{Tianshu Shen}, \bibinfo{person}{Jiaru Li}, \bibinfo{person}{Mohamed~Reda Bouadjenek}, \bibinfo{person}{Zheda Mai}, {and} \bibinfo{person}{Scott Sanner}.} \bibinfo{year}{2023}\natexlab{}.
\newblock \showarticletitle{Towards understanding and mitigating unintended biases in language model-driven conversational recommendation}.
\newblock \bibinfo{journal}{\emph{Information Processing \& Management}} \bibinfo{volume}{60}, \bibinfo{number}{1} (\bibinfo{year}{2023}), \bibinfo{pages}{103139}.
\newblock


\bibitem[S{\"u}hr et~al\mbox{.}(2021)]%
        {suhr2021does}
\bibfield{author}{\bibinfo{person}{Tom S{\"u}hr}, \bibinfo{person}{Sophie Hilgard}, {and} \bibinfo{person}{Himabindu Lakkaraju}.} \bibinfo{year}{2021}\natexlab{}.
\newblock \showarticletitle{Does fair ranking improve minority outcomes? understanding the interplay of human and algorithmic biases in online hiring}. In \bibinfo{booktitle}{\emph{Proceedings of the 2021 AAAI/ACM Conference on AI, Ethics, and Society}}. \bibinfo{pages}{989--999}.
\newblock


\bibitem[Vu et~al\mbox{.}(2021)]%
        {vu2021spot}
\bibfield{author}{\bibinfo{person}{Tu Vu}, \bibinfo{person}{Brian Lester}, \bibinfo{person}{Noah Constant}, \bibinfo{person}{Rami Al-Rfou}, {and} \bibinfo{person}{Daniel Cer}.} \bibinfo{year}{2021}\natexlab{}.
\newblock \showarticletitle{Spot: Better frozen model adaptation through soft prompt transfer}.
\newblock \bibinfo{journal}{\emph{arXiv preprint arXiv:2110.07904}} (\bibinfo{year}{2021}).
\newblock


\bibitem[Wan et~al\mbox{.}(2020)]%
        {wan2020addressing}
\bibfield{author}{\bibinfo{person}{Mengting Wan}, \bibinfo{person}{Jianmo Ni}, \bibinfo{person}{Rishabh Misra}, {and} \bibinfo{person}{Julian McAuley}.} \bibinfo{year}{2020}\natexlab{}.
\newblock \showarticletitle{Addressing marketing bias in product recommendations}. In \bibinfo{booktitle}{\emph{Proceedings of the 13th international conference on web search and data mining}}. \bibinfo{pages}{618--626}.
\newblock


\bibitem[Weydemann et~al\mbox{.}(2019)]%
        {weydemann2019defining}
\bibfield{author}{\bibinfo{person}{Leonard Weydemann}, \bibinfo{person}{Dimitris Sacharidis}, {and} \bibinfo{person}{Hannes Werthner}.} \bibinfo{year}{2019}\natexlab{}.
\newblock \showarticletitle{Defining and measuring fairness in location recommendations}. In \bibinfo{booktitle}{\emph{Proceedings of the 3rd ACM SIGSPATIAL international workshop on location-based recommendations, geosocial networks and geoadvertising}}. \bibinfo{pages}{1--8}.
\newblock


\bibitem[Wu et~al\mbox{.}(2021b)]%
        {wu2021fairness}
\bibfield{author}{\bibinfo{person}{Chuhan Wu}, \bibinfo{person}{Fangzhao Wu}, \bibinfo{person}{Xiting Wang}, \bibinfo{person}{Yongfeng Huang}, {and} \bibinfo{person}{Xing Xie}.} \bibinfo{year}{2021}\natexlab{b}.
\newblock \showarticletitle{Fairness-aware news recommendation with decomposed adversarial learning}. In \bibinfo{booktitle}{\emph{Proceedings of the AAAI Conference on Artificial Intelligence}}, Vol.~\bibinfo{volume}{35}. \bibinfo{pages}{4462--4469}.
\newblock


\bibitem[Wu et~al\mbox{.}(2023)]%
        {wu2023bloomberggpt}
\bibfield{author}{\bibinfo{person}{Shijie Wu}, \bibinfo{person}{Ozan Irsoy}, \bibinfo{person}{Steven Lu}, \bibinfo{person}{Vadim Dabravolski}, \bibinfo{person}{Mark Dredze}, \bibinfo{person}{Sebastian Gehrmann}, \bibinfo{person}{Prabhanjan Kambadur}, \bibinfo{person}{David Rosenberg}, {and} \bibinfo{person}{Gideon Mann}.} \bibinfo{year}{2023}\natexlab{}.
\newblock \showarticletitle{Bloomberggpt: A large language model for finance}.
\newblock \bibinfo{journal}{\emph{arXiv preprint arXiv:2303.17564}} (\bibinfo{year}{2023}).
\newblock


\bibitem[Wu et~al\mbox{.}(2021a)]%
        {wu2021tfrom}
\bibfield{author}{\bibinfo{person}{Yao Wu}, \bibinfo{person}{Jian Cao}, \bibinfo{person}{Guandong Xu}, {and} \bibinfo{person}{Yudong Tan}.} \bibinfo{year}{2021}\natexlab{a}.
\newblock \showarticletitle{TFROM: A Two-sided Fairness-Aware Recommendation Model for Both Customers and Providers}.
\newblock \bibinfo{journal}{\emph{arXiv preprint arXiv:2104.09024}} (\bibinfo{year}{2021}).
\newblock


\bibitem[Xiao et~al\mbox{.}(2020)]%
        {xiao2020enhanced}
\bibfield{author}{\bibinfo{person}{Yang Xiao}, \bibinfo{person}{Qingqi Pei}, \bibinfo{person}{Lina Yao}, \bibinfo{person}{Shui Yu}, \bibinfo{person}{Lei Bai}, {and} \bibinfo{person}{Xianzhi Wang}.} \bibinfo{year}{2020}\natexlab{}.
\newblock \showarticletitle{An enhanced probabilistic fairness-aware group recommendation by incorporating social activeness}.
\newblock \bibinfo{journal}{\emph{Journal of Network and Computer Applications}}  \bibinfo{volume}{156} (\bibinfo{year}{2020}), \bibinfo{pages}{102579}.
\newblock


\bibitem[Xu et~al\mbox{.}(2024)]%
        {xu2024prompting}
\bibfield{author}{\bibinfo{person}{Lanling Xu}, \bibinfo{person}{Junjie Zhang}, \bibinfo{person}{Bingqian Li}, \bibinfo{person}{Jinpeng Wang}, \bibinfo{person}{Mingchen Cai}, \bibinfo{person}{Wayne~Xin Zhao}, {and} \bibinfo{person}{Ji-Rong Wen}.} \bibinfo{year}{2024}\natexlab{}.
\newblock \showarticletitle{Prompting Large Language Models for Recommender Systems: A Comprehensive Framework and Empirical Analysis}.
\newblock \bibinfo{journal}{\emph{arXiv preprint arXiv:2401.04997}} (\bibinfo{year}{2024}).
\newblock


\bibitem[Xu et~al\mbox{.}(2023)]%
        {xu2023openp5}
\bibfield{author}{\bibinfo{person}{Shuyuan Xu}, \bibinfo{person}{Wenyue Hua}, {and} \bibinfo{person}{Yongfeng Zhang}.} \bibinfo{year}{2023}\natexlab{}.
\newblock \showarticletitle{OpenP5: Benchmarking Foundation Models for Recommendation}.
\newblock \bibinfo{journal}{\emph{arXiv preprint arXiv:2306.11134}} (\bibinfo{year}{2023}).
\newblock


\bibitem[Yang et~al\mbox{.}(2020)]%
        {yang2020fairness}
\bibfield{author}{\bibinfo{person}{Forest Yang}, \bibinfo{person}{Mouhamadou Cisse}, {and} \bibinfo{person}{Sanmi Koyejo}.} \bibinfo{year}{2020}\natexlab{}.
\newblock \showarticletitle{Fairness with overlapping groups; a probabilistic perspective}.
\newblock \bibinfo{journal}{\emph{Advances in neural information processing systems}}  \bibinfo{volume}{33} (\bibinfo{year}{2020}), \bibinfo{pages}{4067--4078}.
\newblock


\bibitem[Zhang et~al\mbox{.}(2023)]%
        {zhang2023chatgpt}
\bibfield{author}{\bibinfo{person}{Jizhi Zhang}, \bibinfo{person}{Keqin Bao}, \bibinfo{person}{Yang Zhang}, \bibinfo{person}{Wenjie Wang}, \bibinfo{person}{Fuli Feng}, {and} \bibinfo{person}{Xiangnan He}.} \bibinfo{year}{2023}\natexlab{}.
\newblock \showarticletitle{Is chatgpt fair for recommendation? evaluating fairness in large language model recommendation}. In \bibinfo{booktitle}{\emph{Proceedings of the 17th ACM Conference on Recommender Systems}}. \bibinfo{pages}{993--999}.
\newblock


\bibitem[Zhang et~al\mbox{.}(2024)]%
        {zhang2024understanding}
\bibfield{author}{\bibinfo{person}{Lemei Zhang}, \bibinfo{person}{Peng Liu}, \bibinfo{person}{Yashar Deldjoo}, \bibinfo{person}{Yong Zheng}, {and} \bibinfo{person}{Jon~Atle Gulla}.} \bibinfo{year}{2024}\natexlab{}.
\newblock \showarticletitle{Understanding Language Modeling Paradigm Adaptations in Recommender Systems: Lessons Learned and Open Challenges}. In \bibinfo{booktitle}{\emph{The 27th European Conference on Artificial Intelligence (ECAI'24)}}.
\newblock


\bibitem[Zhu et~al\mbox{.}(2021)]%
        {zhu2021fairness}
\bibfield{author}{\bibinfo{person}{Ziwei Zhu}, \bibinfo{person}{Jingu Kim}, \bibinfo{person}{Trung Nguyen}, \bibinfo{person}{Aish Fenton}, {and} \bibinfo{person}{James Caverlee}.} \bibinfo{year}{2021}\natexlab{}.
\newblock \showarticletitle{Fairness among new items in cold start recommender systems}. In \bibinfo{booktitle}{\emph{Proceedings of the 44th International ACM SIGIR Conference on Research and Development in Information Retrieval}}. \bibinfo{pages}{767--776}.
\newblock


\end{thebibliography}


\end{document}